\documentclass[journal]{IEEEtran}
\usepackage{cite}

\usepackage{amssymb,amsfonts}

\usepackage{lipsum}
\usepackage{mathtools}
\usepackage{cuted}

\usepackage{balance}

\usepackage{amsmath}

\newtheorem{theorem}{\it Theorem}

\newtheorem{corollary}{\it Corollary}
\newtheorem{proposition}{\it Proposition}
\newtheorem{example}{\it Example}
\newtheorem{observation}{\it Observation}

\ifCLASSINFOpdf
\else
\fi
\begin{document}
%
\title{Feedback Capacity and a Variant of the Kalman Filter with ARMA Gaussian Noises: Explicit Bounds and Feedback Coding Design}
%
%
%

\author{Song~Fang~and~Quanyan~Zhu
\thanks{Song Fang and Quanyan Zhu are with the Department
of Electrical and Computer Engineering, New York University, USA (e-mail: song.fang@nyu.edu; quanyan.zhu@nyu.edu).}
\thanks{This paper was presented in part at IEEE ISIT 2020 as \cite{fang2020connection}, which, however, only considered the class of scalar plant parameters as $A = a \in \mathbb{R}$ and $C = c \in \mathbb{R}$; in other words, the main result therein is essentially Corollary~\ref{n1} herein, a special case of what are presented in this paper. Note also that all the proofs in \cite{fang2020connection} were omitted due to lack of space.}
}

\maketitle

\begin{abstract}
In this paper, we relate a feedback channel with any finite-order autoregressive moving-average (ARMA) Gaussian noises to a variant of the Kalman filter. In light of this, we obtain relatively explicit lower bounds on the feedback capacity for such colored Gaussian noises, and the bounds are seen to be consistent with various existing results in the literature. Meanwhile, this variant of the Kalman filter also leads to explicit recursive coding schemes with clear structures to achieve the lower bounds. In general, our results provide an alternative perspective while pointing to potentially tighter bounds for the feedback capacity problem.
\end{abstract}


%
\IEEEpeerreviewmaketitle

\section{Introduction}
%
%
%
%
\IEEEPARstart{T}{he}
feedback capacity \cite{cover1989gaussian} of additive colored Gaussian noise channels has been a long-standing problem in information theory, generating numerous research results over the years, due to its interdisciplinary nature and significance in understanding and applying communication/coding with feedback.
In general, we refer to the breakthrough paper \cite{kim2010feedback} and the references therein for a rather complete literature review; see also \cite{kim2006feedback, kim2006gaussian} for possibly complementary paper surveys. Meanwhile, papers on this topic have also been coming out continuously after \cite{kim2010feedback}, which include but are certainly not restricted to \cite{ardestanizadeh2012control, liu2014convergence,  stavrou2017sequential,
	liu2018feedback,
	li2018youla, rawat2018computation, pedram2018some,
	kourtellaris2018information, gattami2018feedback, ihara2019feedback, fang2020connection, aharoni2020capacity}. More specifically, in \cite{kim2010feedback}, Kim proposed a variational characterization of the feedback capacity in terms of the noise power spectral density and showed that the feedback capacity can be achieved by linear feedback coding schemes, which, among the many other conclusions derived therein, yields for the first time an analytic
expression when
specialized to the first-order autoregressive moving-average (ARMA) noises.
In \cite{ardestanizadeh2012control}, Ardestanizadeh and  Franceschetti showed, from the perspective of Bode integral, that the feedback capacity is equal to the maximum instability that can be tolerated by any linear controller under a given power constraint. In \cite{liu2014convergence}, Liu and Elia established the mutual equivalences among the feedback capacity, the Cram\'er-Rao bound or the minimum mean-square error in estimation systems, and the Bode integral in control systems. In \cite{stavrou2017sequential}, Stavrou {\em et al.} obtained sequential necessary and sufficient
conditions in terms of input conditional distributions to achieve the feedback capacity.
In \cite{liu2018feedback}, Liu and Han proved the uniqueness of the optimal solution to the variational characterization of \cite{kim2010feedback} and proposed
an algorithm to recursively compute the optimal solution with convergence guarantee, while, for any finite-order ARMA noises, providing a relatively more explicit formula as a
simple function evaluated at a solution to a system of polynomial equations.
In \cite{li2018youla}, Li and Elia showed that the problem of 
achieving feedback capacity coincides with the problem of finding stabilizing feedback controllers
with maximal transmission rate over Youla parameters, and proposed an approach to numerically compute the feedback capacity while constructing feedback codes that are arbitrarily close to capacity-achieving.
Subsequently in \cite{rawat2018computation}, Rawat {\em et al.} generalized the approach in \cite{li2018youla} to the feedback capacity of multi-antenna channels with multivariate colored noises.
In \cite{pedram2018some}, Pedram and Tanaka analyzed structural properties of the
optimal feedback policies and developed a convex program
that can be used to compute the feedback capacity. 
In \cite{kourtellaris2018information}, Kourtellaris and Charalambous studied the feedback capacity problem by applying stochastic optimal control theory and a variational equality of directed information while developing a methodology to
identify the information structures of optimal channel input conditional distributions.
In \cite{gattami2018feedback}, Gattami introduced a new approach to the feedback capacity problem by solving the problem over a finite number of transmissions
and then taking the limit of an infinite number of transmissions. 
In \cite{ihara2019feedback}, Ihara presented an alternative proof to the analytic expression for the first-order moving average (MA) noises. In \cite{aharoni2020capacity}, Aharoni {\em et al.} proposed a directed information estimation algorithm based on neural networks to compute the feedback capacity.

In particular, analytic or relatively explicit expressions or lower bounds, oftentimes in terms of a root of a polynomial equation, of the feedback capacity have been presented in \cite{butman1969general, butman1976linear, Eli:04, yang2007feedback, kim2006feedback, kim2010feedback, liu2018feedback, ihara2019feedback}. More specifically, Butman obtained explicit lower bounds for the first-order autoregressive (AR) noises in
\cite{butman1969general} and then for any finite-order AR noises in \cite{butman1976linear}. In \cite{Eli:04}, Elia derived a refined, explicit lower bound for any finite-order AR noises.
In \cite{yang2007feedback}, Yang {\em et al.} obtained an explicit lower bound for the first-order ARMA noises.
For the first time, Kim discovered analytic formulae of feedback capacity, rather than its lower bounds, for the first-order AR noises in \cite{kim2006feedback} and then for the first-order ARMA noises in \cite{kim2010feedback}. 
In \cite{liu2018feedback}, Liu and Han discovered a relatively explicit expression for any finite-order ARMA noise in terms of a solution to a system of polynomial equations. 
In \cite{ihara2019feedback}, Ihara presented an alternative proof to the analytic expression for the first-order MA noises derived in \cite{kim2006feedback}. This line of work provided the main motivations for obtaining the results in this paper.

In general, this paper employs a control-theoretic approach to analyze feedback channels, which has been inspired by, e.g., \cite{Eli:04, ardestanizadeh2012control, liu2014convergence} (Bode integral), \cite{yang2007feedback,  tatikonda2008capacity, liu2014convergence,
	pedram2018some, kourtellaris2018information, gattami2018feedback} (stochastic control and/or Kalman filter), as well as \cite{
	li2018youla, rawat2018computation} (Youla parametrization), in a broad sense; see also \cite{fang2017towards} and the references therein. 
One difference
from the previous works, however, is that in this paper we
adopt a particular variant of the Kalman filter that can deal with the first-order AR noises
without extending the state to be estimated in Kalman filtering
systems, as introduced in \cite{anderson2012optimal}, and generalize the approach to
cope with any finite-order ARMA noises. Another difference is that we examine the algebraic Riccati equation associated with the Kalman filter in a ``non-recursive" characterization recently developed in \cite{FangACC18}.
Accordingly,
we establish the connection between this variant of the Kalman filter and a feedback channel with any finite-order ARMA noises,
after carrying out a series of equivalent transformations. 
In light of this connection, we obtain explicit lower bounds on the feedback capacity for any finite-order ARMA noises, by designing the parameters of the plant in the Kalman filtering system in a structural way. In addition, this variant of the Kalman filter naturally provides explicit recursive coding schemes with clear structures to achieve the lower bounds.

The lower bounds presented in this paper are seen to be consistent with various existing (analytic or explicit) results in the literature. Specifically, our bounds are shown to be
tight for the first-order ARMA noises \cite{kim2010feedback} (see also \cite{liu2018feedback}) and a special class of the second-order MA noises \cite{kim2006feedback}. Meanwhile, our results are seen to reduce to the lower bounds
for any finite-order AR noises of \cite{Eli:04} (see also \cite{butman1976linear}).
Particularly, in comparison to the explicit expression of feedback capacity for any finite-order ARMA noises derived in \cite{liu2018feedback} (see Theorem~18 therein), 1) what we obtained are lower bounds; 2) our explicit expression admits a simpler form; 3) it is not yet fully clear what the relationship between the two is. (Note that the approach taken in \cite{liu2018feedback} was not from the perspective of Kalman filters.) In general, our results shall complement those of \cite{liu2018feedback} in the sense that examining the gap between the expressions of \cite{liu2018feedback} and ours might likely either simplify the expression in \cite{liu2018feedback} or point to structures of the plant parameters with tighter bounds in our approach; in either case, additional insights will be gained into the feedback capacity problem. On the other hand, the explicit recursive coding schemes developed
in this paper also complement the existing ones in the literature (see, e.g., \cite{Eli:04, kim2010feedback}). 






The remainder of the paper is organized as follows. Section~II provides technical preliminaries. Section~III presents the main results. Concluding remarks are given in Section~IV.

\section{Preliminaries}

In this paper, we consider real-valued continuous random variables and discrete-time stochastic processes they compose. All random variables and stochastic processes are assumed to be zero-mean for simplicity and without loss of generality. We represent random variables using boldface letters. The logarithm is defined with base $2$.
A stochastic process $\left\{ \mathbf{x}_{k}\right\}$ is said to be asymptotically stationary if it is stationary as $k \to \infty$, and herein stationarity means strict stationarity \cite{Pap:02}. Note in particular that, for simplicity and
with abuse of notations, we utilize $\mathbf{x} \in \mathbb{R}$ and $\mathbf{x} \in \mathbb{R}^n$ to
indicate that $\mathbf{x}$ is a real-valued random variable and that $\mathbf{x}$
is a real-valued $n$-dimensional random vector, respectively. Definitions and properties of the information-theoretic notions
such as entropy rate $h_\infty \left(\mathbf{x}\right)$ can be found in, e.g., \cite{Cov:06}.

%

\subsection{Feedback Capacity} \label{notation}

Consider an additive colored Gaussian noise channel with feedback given as 
\begin{flalign}\mathbf{y}_{k} = \mathbf{x}_{k} \left( \mathbf{m}, \mathbf{y}_{1, \ldots, k-1} \right) + \mathbf{z}_{k},~k=1, 2, \ldots \nonumber
\end{flalign}
where $\left\{ \mathbf{x}_{k} \right\}, \mathbf{x}_{k} \in \mathbb{R}$ denotes the channel input, $\left\{ \mathbf{y}_{k} \right\}, \mathbf{y}_{k} \in \mathbb{R}$ denotes the channel output, $\left\{ \mathbf{z}_{k} \right\}, \mathbf{z}_{k} \in \mathbb{R}$ denotes the additive noise that is assumed to be stationary colored Gaussian, and $\mathbf{m}$ denotes the message. The feedback capacity $C_{\text{f}}$ of such a channel with power constraint $\overline{P}$ is defined as \cite{cover1989gaussian} 
\begin{flalign} \label{fcdef1}
C_{\text{f}} 
&=  \lim_{k\to \infty} \sup_{ \mathbf{x}_{1, \ldots, k} } \frac{ I \left( \mathbf{m} ; \mathbf{y}_{1, \ldots, k} \right)}{k} \nonumber \\
&=  \lim_{k\to \infty} \sup_{ \mathbf{x}_{1, \ldots, k} } \frac{ h \left( \mathbf{y}_{1, \ldots, k} \right) - h \left( \mathbf{z}_{1, \ldots, k} \right)}{k}, 
\end{flalign}
where the supremum is taken over all channel input $\mathbf{x}_{1, \ldots, k}$ that satisfy
\begin{flalign}
\frac{1}{k} \sum_{i=1}^{k} \mathbb{E} \left[ \mathbf{x}^2_{i} \left( \mathbf{m}, \mathbf{y}_{1, \ldots, i-1} \right) \right] \leq \overline{P}. \nonumber
\end{flalign}
Recently in \cite{kim2010feedback}, it was discovered that \eqref{fcdef1} is equal to
\begin{flalign} \label{fcdef}
C_{\text{f}} = \sup_{\left\{ \mathbf{x}_{k} \right\}} \left[ h_{\infty} \left( \mathbf{y} \right) -  h_{\infty} \left(  \mathbf{z} \right) \right],
\end{flalign}
where the supremum is taken over all stationary channel input processes $\left\{ \mathbf{x}_{k} \right\}$ of the form 
\begin{flalign} \label{fcdef2}
\mathbf{x}_{k} = \sum_{i=1}^{\infty} b_{i} \mathbf{z}_{k-i},~b_{i} \in \mathbb{R}, 
\end{flalign} 
while satisfying 
\begin{flalign}
\mathbb{E} \left[ \mathbf{x}^2_{k} \right] \leq \overline{P}. \nonumber
\end{flalign}
In fact, \cite{kim2010feedback} provided a sequence of equivalent characterizations of the feedback capacity, whereas for the purpose of this paper, it suffices to adopt the characterization of \eqref{fcdef} and \eqref{fcdef2} herein (see the proof of Theorem~\ref{fc4}).

\subsection{Kalman Filter} \label{kalamnsection}


\begin{figure}
	\begin{center}
		\vspace{-3mm}
		\includegraphics [width=0.5\textwidth]{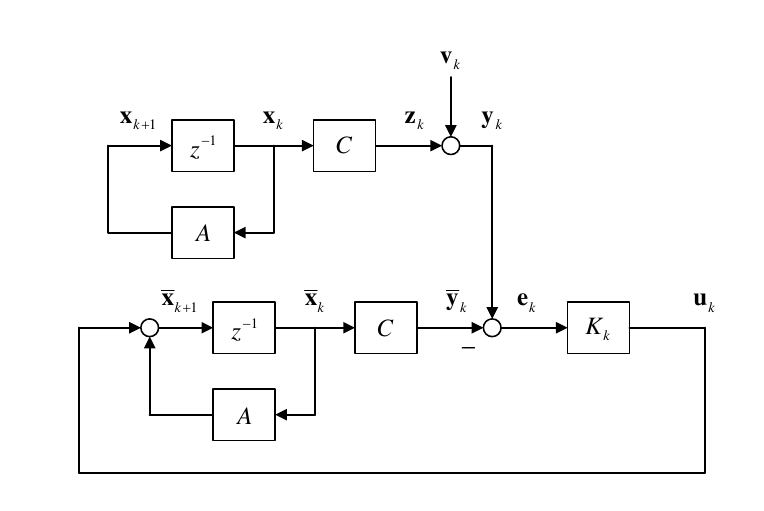}
		\vspace{-6mm}
		\caption{The Kalman filtering system.}
		\label{kalman1}
	\end{center}
	\vspace{-3mm}
\end{figure}

We now give a brief review of (a special case of) the Kalman filter \cite{linearestimation, anderson2012optimal}; note that hereinafter the notations are not to
be confused with those in Section~\ref{notation}. Particularly, consider the Kalman filtering system depicted in Fig.~\ref{kalman1}, where the state-space model of the plant to be estimated is given by
\begin{flalign} \label{plant}
\left\{ \begin{array}{rcl}
\mathbf{x}_{k+1} & = & A \mathbf{x}_{k},\\
\mathbf{y}_{k} & = & C \mathbf{x}_{k} + \mathbf{v}_k.
\end{array} \right. 
\end{flalign}
Herein, $\mathbf{x}_{k} \in \mathbb{R}^n$ is the state to be estimated, $\mathbf{y}_{k} \in \mathbb{R}$ is the plant output, and $\mathbf{v}_{k} \in \mathbb{R}$ is the measurement noise, whereas the process noise, normally denoted as $\left\{ \mathbf{w}_{k} \right\}$ \cite{linearestimation, anderson2012optimal}, is assumed to be absent. The system matrix is $ A \in \mathbb{R}^{n \times n}$ while the output matrix is $ C \in \mathbb{R}^{1 \times n}$, and we assume that $A$ is anti-stable (i.e., all the eigenvalues are unstable with magnitude greater than or equal to $1$) while the pair $\left( A, C \right)$ is observable (and thus detectable \cite{astrom2010feedback}). Suppose that $\left\{ \mathbf{v}_{k} \right\}$ is white Gaussian with variance $\sigma_{\mathbf{v}}^2 = \mathbb{E} \left[ \mathbf{v}_{k}^{2} \right] \geq  0$ and the initial state $\mathbf{x}_{0}$ is Gaussian with covariance $ \mathbb{E} \left[ \mathbf{x}_0 \mathbf{x}_0^{\mathrm{T}} \right] \succ 0$. Furthermore, $\left\{ \mathbf{v}_{k} \right\}$ and $\mathbf{x}_{0}$ are assumed to be uncorrelated.  
Correspondingly, the Kalman filter (in the observer form \cite{astrom2010feedback}) for \eqref{plant} is given by
\begin{flalign} \label{estimator}
\left\{ \begin{array}{rcl}
\overline{\mathbf{x}}_{k+1}&=&A \overline{\mathbf{x}}_{k} +\mathbf{u}_k, \\
\overline{\mathbf{y}}_{k}&= & C \overline{\mathbf{x}}_{k}, \\
\mathbf{e}_{k}&=&\mathbf{y}_{k}-\overline{\mathbf{y}}_k, \\
\mathbf{u}_{k}&=& K_{k} \mathbf{e}_{k},
\end{array} 
\right.
\end{flalign}
where $\overline{\mathbf{x}}_{k} \in \mathbb{R}^n$, $\overline{\mathbf{y}}_{k} \in \mathbb{R}$, $\mathbf{e}_{k} \in \mathbb{R}$, and $\mathbf{\mathbf{u}}_{k} \in \mathbb{R}^n$. Herein, $K_{k}$ denotes the observer gain \cite{astrom2010feedback} (note that the observer gain is different from the Kalman gain by a factor of $A$; see, e.g.,  \cite{anderson2012optimal, astrom2010feedback} for more details) given by
\begin{flalign}
K_{k} = A P_{k} C^{\mathrm{T}} \left( C P_{k} C^{\mathrm{T}} + \sigma_{\mathbf{v}}^2 \right)^{-1}, \nonumber
\end{flalign}
where $P_k$ denotes the state estimation error covariance as
\begin{flalign}
P_k = \mathbb{E} \left[ \left(\mathbf{x}_k -\overline{\mathbf{x}}_k \right) \left(\mathbf{x}_k -\overline{\mathbf{x}}_k \right)^{\mathrm{T}} \right]. \nonumber
\end{flalign}
In addition, $P_k$ can be obtained iteratively by the Riccati equation
\begin{flalign}
P_{k+1}
= A P_{k} A^{\mathrm{T}} - A P_{k} C^{\mathrm{T}} \left( C P_{k} C^{\mathrm{T}} + \sigma_{\mathbf{v}}^2 \right)^{-1} C P_{k} A^{\mathrm{T}}, \nonumber
\end{flalign}
with $P_{0} \succ 0$.
Additionally, it is known \cite{linearestimation, anderson2012optimal} that since $\left( A, C \right)$ is detectable, the Kalman filtering system converges, i.e., the state estimation error $\left\{ \mathbf{x}_k -\overline{\mathbf{x}}_k \right\}$ is asymptotically stationary. Moreover, in steady state, the optimal state estimation error variance
\begin{flalign}
P =\lim_{k\to \infty} \mathbb{E} \left[ \left(\mathbf{x}_k -\overline{\mathbf{x}}_k \right) \left(\mathbf{x}_k -\overline{\mathbf{x}}_k \right)^{\mathrm{T}} \right] \nonumber
\end{flalign} 
attained by the Kalman filter is given by the (non-zero) positive semi-definite solution \cite{anderson2012optimal} to the algebraic Riccati equation
\begin{flalign} 
P
= A P A^{\mathrm{T}} - A P C^{\mathrm{T}} \left( C P C^{\mathrm{T}} + \sigma_{\mathbf{v}}^2 \right)^{-1} C P A^{\mathrm{T}}, \nonumber
\end{flalign} 
whereas the steady-state observer gain is given by 
\begin{flalign}  
K = A P C^{\mathrm{T}} \left( C P C^{\mathrm{T}} + \sigma_{\mathbf{v}}^2 \right)^{-1}. \nonumber
\end{flalign}  

Meanwhile, it is known from \cite{FangACC18} (by letting $m = 1$ and $W = 0$ in Theorem~1 therein; since the fact that $\left\{ \mathbf{w}_{k} \right\}$ is absent implicates $W = \mathbb{E} \left[ \mathbf{w}_k \mathbf{w}_k^{\mathrm{T}} \right] = 0$) that
\begin{flalign}  \label{bode}
C P C^{\mathrm{T}} + \sigma_{\mathbf{v}}^2 
&= \left[ \prod_{\ell=1}^{n} \max \left\{ 1, \left| \lambda_{\ell} \right|^2 \right\} \right] \sigma_{\mathbf{v}}^2 \nonumber \\
& = \left( \prod_{\ell=1}^{n} \left| \lambda_{\ell} \right|^2 \right) \sigma_{\mathbf{v}}^2 
= \left( \prod_{\ell=1}^{n} \lambda_{\ell}^2 \right) \sigma_{\mathbf{v}}^2,
\end{flalign}
where
\begin{flalign} 
\prod_{\ell=1}^{n} \max \left\{ 1, \left| \lambda_{\ell} \right|^2 \right\} = \prod_{\ell=1}^{n} \left| \lambda_{\ell} \right|^2,
\end{flalign}
since $A$ is assumed to be anti-stable, whereas 
\begin{flalign} 
\prod_{\ell=1}^{n} \left| \lambda_{\ell} \right|^2
= \prod_{\ell=1}^{n} \lambda_{\ell}^2,
\end{flalign}
since $A$ is a real matrix.

\begin{figure}
	\begin{center}
		\vspace{-3mm}
		\includegraphics [width=0.5\textwidth]{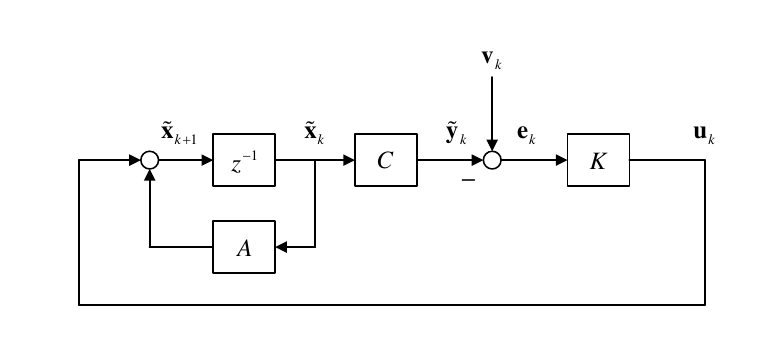}
		\vspace{-6mm}
		\caption{The steady-state Kalman filtering system in integrated form.}
		\label{kalman2}
	\end{center}
	\vspace{-3mm}
\end{figure}

In fact, by letting $\widetilde{\mathbf{x}}_{k} = \overline{\mathbf{x}}_{k} - \mathbf{x}_{k} $ and $\widetilde{\mathbf{y}}_{k} = \overline{\mathbf{y}}_{k} - \mathbf{z}_{k} = \overline{\mathbf{y}}_{k} - C \mathbf{x}_{k} $, we may integrate the systems of \eqref{plant} and \eqref{estimator} in steady state into an equivalent form:
\begin{flalign}
\left\{ \begin{array}{rcl}
\widetilde{\mathbf{x}}_{k+1}&=&A \widetilde{\mathbf{x}}_{k} +\mathbf{u}_k, \\
\widetilde{\mathbf{y}}_{k}&= & C \widetilde{\mathbf{x}}_{k}, \\
\mathbf{e}_{k}&=& - \widetilde{\mathbf{y}}_k + \mathbf{v}_k, \\
\mathbf{u}_{k}&=& K \mathbf{e}_{k},
\end{array}
\right. 
\end{flalign}
as depicted in Fig.~\ref{kalman2}, since all the sub-systems are linear. 

\section{Feedback Capacity for ARMA Gaussian Noises} \label{sectionAR}

%

The approach we take in this paper to obtain lower bounds on the feedback capacity of channels with ARMA Gaussian noises is by establishing the connection between such feedback channels and a variant of the Kalman filter to deal with ARMA Gaussian noises. Towards this end, we first present the following variant of the Kalman filter.

\subsection{A Variant of the Kalman Filter}


Consider again the Kalman filtering system given in Fig.~\ref{kalman1}. Suppose that the plant to be estimated is still given by
\begin{flalign} \label{plant33}
\left\{ \begin{array}{rcl}
\mathbf{x}_{k+1} & = & A \mathbf{x}_{k},\\
\mathbf{y}_{k} & = & C \mathbf{x}_{k} +\mathbf{v}_k,
\end{array} \right.
\end{flalign} 
only this time with an ARMA measurement noise $\left\{ \mathbf{v}_{k} \right\}, \mathbf{v}_{k} \in \mathbb{R}$, represented as
\begin{flalign} \label{ARMA}
\mathbf{v}_{k} 
&= \sum_{i=1}^{p} f_{i} \mathbf{v}_{k-i} + \widehat{\mathbf{v}}_k + \sum_{j=1}^{q} g_{j} \widehat{\mathbf{v}}_{k-j} 
, 
\end{flalign}
where $\left\{ \widehat{\mathbf{v}}_k \right\}, \widehat{\mathbf{v}}_k \in \mathbb{R}$ is white Gaussian with variance $\sigma_{\widehat{\mathbf{v}}}^2 = \mathrm{E} \left[ \widehat{\mathbf{v}}_{k}^2 \right] > 0$.
Equivalently, $\left\{ \mathbf{v}_k \right\}$ may be represented \cite{vaidyanathan2007theory} as the output of a linear time-invariant (LTI) filter $F \left( z \right)$ driven by input $\left\{ \widehat{\mathbf{v}}_k \right\}$, where
\begin{flalign}
F \left( z \right) 
= \frac{1 + \sum_{j=1}^{q} g_{j} z^{-j}}{1 - \sum_{i=1}^{p} f_{i} z^{-i}}
.
\end{flalign}
Herein, we assume that $F \left( z \right)$ is stable and minimum-phase.

We may then employ the method of dealing with colored noises without extending the state of the Kalman filter, as introduced in \cite{anderson2012optimal} (Chapter~11), after certain modifications. In fact, since therein the process noise $\left\{ \mathbf{w}_{k} \right\}$ is not absent, this approach is only applicable to the first-order AR noises; whereas in this paper, assuming that the process noise is absent, we may generalize the approach to any finite-order ARMA noises.


\begin{proposition} \label{anderson11}
	Denote
	\begin{flalign}
	\widehat{\mathbf{y}}_{k} 
	= - \sum_{j=1}^{q} g_{j} \widehat{\mathbf{y}}_{k-j} + \mathbf{y}_{k} - \sum_{i=1}^{p} f_{i} \mathbf{y}_{k-i}.
	\end{flalign}
	Then, \eqref{plant33} is equivalent to
	\begin{flalign} \label{plant22}
	\left\{ \begin{array}{rcl}
	\mathbf{x}_{k+1} & = & A \mathbf{x}_{k},\\
	\widehat{\mathbf{y}}_{k}  & = & \widehat{C} \mathbf{x}_{k} + \widehat{\mathbf{v}}_k,
	\end{array} \right.
	\end{flalign}
	where
	\begin{flalign} \label{c33}
	\widehat{C} 
	&= C \left( I - \sum_{i=1}^{p} f_{i} A^{-i} \right) \left( I + \sum_{j=1}^{q} g_{j} A^{-j} \right)^{-1}. 
	\end{flalign}
\end{proposition}

\begin{IEEEproof}
		Note first that since $F \left( z \right)$ is stable and minimum-phase, 
	the inverse filter
	\begin{flalign} \label{inverse}
	F^{-1} \left( z \right)
	=  \frac{1 - \sum_{i=1}^{p} f_{i} z^{-i}}{1 + \sum_{j=1}^{q} g_{j} z^{-j}}
	\end{flalign} 
	is also stable and minimum-phase. As a result, it holds $\forall \left| z \right| \geq 1$ that
	\begin{flalign}
	\frac{1 - \sum_{i=1}^{p} f_{i} z^{-i}}{1 + \sum_{j=1}^{q} g_{j} z^{-j}} \neq 0,\nonumber
	\end{flalign}
	i.e., the region of convergence must include, though not necessarily restricted to, $\left| z \right| \geq 1$.
	Consequently, for $\left| z \right| \geq 1$, we may expand 
	\begin{flalign}
	\frac{1 - \sum_{i=1}^{p} f_{i} z^{-i}}{1 + \sum_{j=1}^{q} g_{j} z^{-j}} = 1 - \sum_{i=1}^{\infty} h_{i} z^{-i}
	, \nonumber 
	\end{flalign}
	and thus $\left\{ \widehat{\mathbf{v}}_k \right\}$ can be reconstructed from $\left\{ \mathbf{v}_k \right\}$ as \cite{vaidyanathan2007theory} 
	\begin{flalign} 
	\widehat{\mathbf{v}}_{k}
	=  \mathbf{v}_{k} - \sum_{i=1}^{\infty} h_{i} \mathbf{v}_{k-i}
	= - \sum_{j=1}^{q} g_{j} \widehat{\mathbf{v}}_{k-j} + \mathbf{v}_{k} - \sum_{i=1}^{p} f_{i} \mathbf{v}_{k-i} 
	.  \nonumber
	\end{flalign}
	Accordingly, it holds that
	\begin{flalign}
	\widehat{\mathbf{y}}_{k} 
	&= 	- \sum_{j=1}^{q} g_{j} \widehat{\mathbf{y}}_{k-j} + \mathbf{y}_{k} - \sum_{i=1}^{p} f_{i} \mathbf{y}_{k-i}
	=  \mathbf{y}_{k} - \sum_{i=1}^{\infty} h_{i} \mathbf{y}_{k-i} \nonumber \\
	&=  \mathbf{y}_{k} - \sum_{i=1}^{\infty} h_{i} \left( C \mathbf{x}_{k-i} +\mathbf{v}_{k-i} \right) \nonumber \\
	&=  C \mathbf{x}_{k} - \sum_{i=1}^{\infty} h_{i} C \mathbf{x}_{k-i} + \mathbf{v}_{k} - \sum_{i=1}^{\infty} h_{i} \mathbf{v}_{k-i} \nonumber \\
	&=  C \mathbf{x}_{k}- \sum_{i=1}^{\infty} h_{i} C \mathbf{x}_{k-i} + \widehat{\mathbf{v}}_k. \nonumber
	\end{flalign}
	Meanwhile, since $A$ is anti-stable (and thus invertible), we have $\mathbf{x}_{k-i} = A^{-i} \mathbf{x}_{k}$. As a result, 
	\begin{flalign}
	\widehat{\mathbf{y}}_{k} 
	&=  C \mathbf{x}_{k} - \sum_{i=1}^{\infty} h_{i} C \mathbf{x}_{k-i} + \widehat{\mathbf{v}}_k \nonumber \\
	&=  C \left( I - \sum_{i=1}^{\infty} h_{i} A^{-i} \right) \mathbf{x}_{k} + \widehat{\mathbf{v}}_k. \nonumber
	\end{flalign}
    Furthermore,
	\begin{flalign}
	&I - \sum_{i=1}^{\infty} h_{i} A^{-i}
	= I - \sum_{i=1}^{\infty} h_{i} \left( T \Lambda T^{-1} \right)^{-i} \nonumber \\
	&~~~~= T \left( I - \sum_{i=1}^{\infty} h_{i} \Lambda^{-i} \right) T^{-1} \nonumber \\
	&~~~~= T \left( I - \sum_{i=1}^{\infty} h_{i} \left[
	\begin{array}{cccc}
	\lambda_{1} & 0 & \cdots & 0\\
	0 & \lambda_{2} & \cdots & 0\\
	\vdots & \vdots & \ddots & \vdots\\
	0 & 0 & \cdots & \lambda_{n}\\
	\end{array} \right]^{-i} \right) T^{-1} \nonumber \\
	&~~~~= T \left[
	\begin{array}{cccc}
	1 - \sum_{i=1}^{\infty} h_{i} \lambda_{1}^{-i} & \cdots & 0\\
	\vdots  & \ddots & \vdots\\
	0 & \cdots & 1 - \sum_{i=1}^{\infty} h_{i} \lambda_{n}^{-i}\\
	\end{array} \right] T^{-1}.
	\nonumber
	\end{flalign}
	On the other hand, we have shown that
	\begin{flalign}
	1 - \sum_{i=1}^{\infty} h_{i} z^{-i} = \frac{1 - \sum_{i=1}^{p} f_{i} z^{-i}}{1 + \sum_{j=1}^{q} g_{j} z^{-j}},
	\nonumber
	\end{flalign}
	i.e., $1 - \sum_{i=1}^{\infty} h_{i} z^{-i}$ converges, for $\left| z \right| \geq 1$.
	As such, since $\left| \lambda_{\ell} \right| \geq 1, \ell = 1, \ldots, n$, we have
	\begin{flalign}
	&\left[
	\begin{array}{cccc}
	1 - \sum_{i=1}^{\infty} h_{i} \lambda_{1}^{-i} & \cdots & 0\\
	\vdots  & \ddots & \vdots\\
	0 & \cdots & 1 - \sum_{i=1}^{\infty} h_{i} \lambda_{n}^{-i}\\
	\end{array} \right] \nonumber \\
	&~~~~= \left[
	\begin{array}{cccc}
	\frac{1 - \sum_{i=1}^{p} f_{i} \lambda_{1}^{-i}}{1 + \sum_{j=1}^{q} g_{j} \lambda_{1}^{-j}} & \cdots & 0\\
	\vdots  & \ddots & \vdots\\
	0 & \cdots & \frac{1 - \sum_{i=1}^{p} f_{i} \lambda_{n}^{-i}}{1 + \sum_{j=1}^{q} g_{j} \lambda_{n}^{-j}}\\
	\end{array} \right] \nonumber \\
	&~~~~= \left[
	\begin{array}{cccc}
	1 - \sum_{i=1}^{p} f_{i} \lambda_{1}^{-i} & \cdots & 0\\
	\vdots  & \ddots & \vdots\\
	0 & \cdots & 1 - \sum_{i=1}^{p} f_{i} \lambda_{n}^{-i}\\
	\end{array} \right]  \nonumber \\
	&~~~~~~~~ \times \left[
	\begin{array}{cccc}
	1 - \sum_{j=1}^{q} g_{j} \lambda_{1}^{-j} & \cdots & 0\\
	\vdots  & \ddots & \vdots\\
	0 & \cdots & 1 - \sum_{j=1}^{q} g_{j} \lambda_{n}^{-j}\\
	\end{array} \right] \nonumber \\
	&~~~~= \left( I - \sum_{i=1}^{p} f_{i} \left[
	\begin{array}{cccc}
	\lambda_{1} & 0 & \cdots & 0\\
	0 & \lambda_{2} & \cdots & 0\\
	\vdots & \vdots & \ddots & \vdots\\
	0 & 0 & \cdots & \lambda_{n}\\
	\end{array} \right]^{-i} \right) \nonumber \\
	&~~~~~~~~ \times \left( I - \sum_{j=1}^{q} g_{j} \left[
	\begin{array}{cccc}
	\lambda_{1} & 0 & \cdots & 0\\
	0 & \lambda_{2} & \cdots & 0\\
	\vdots & \vdots & \ddots & \vdots\\
	0 & 0 & \cdots & \lambda_{n}\\
	\end{array} \right]^{-j} \right) \nonumber \\
	&~~~~= \left( I - \sum_{i=1}^{p} f_{i} \Lambda^{-i} \right) \left( I + \sum_{j=1}^{q} g_{j} \Lambda^{-j} \right)^{-1} 
	,
	\nonumber 
	\end{flalign}
	and hence
	\begin{flalign}
	&T \left[
	\begin{array}{cccc}
	1 - \sum_{i=1}^{\infty} h_{i} \lambda_{1}^{-i} & \cdots & 0\\
	\vdots  & \ddots & \vdots\\
	0 & \cdots & 1 - \sum_{i=1}^{\infty} h_{i} \lambda_{n}^{-i}\\
	\end{array} \right] T^{-1} \nonumber \\
	&~~~~= T \left( I - \sum_{i=1}^{p} f_{i} \Lambda^{-i} \right) \left( I + \sum_{j=1}^{q} g_{j} \Lambda^{-j} \right)^{-1} T^{-1} \nonumber \\
	&~~~~= T \left( I - \sum_{i=1}^{p} f_{i} \Lambda^{-i} \right) T^{-1} T \left( I + \sum_{j=1}^{q} g_{j} \Lambda^{-j} \right)^{-1} T^{-1} \nonumber \\
	&~~~~= \left( I - \sum_{i=1}^{p} f_{i} T \Lambda^{-i} T^{-1} \right) \left( I + \sum_{j=1}^{q} g_{j} T \Lambda^{-j} T^{-1} \right)^{-1}  \nonumber \\
	&~~~~= \left[ I - \sum_{i=1}^{p} f_{i} \left( T \Lambda T^{-1} \right)^{-i} \right] \left[ I + \sum_{j=1}^{q} g_{j} \left( T \Lambda T^{-1} \right)^{-j} \right]^{-1} \nonumber \\
	&~~~~
	=\left( I - \sum_{i=1}^{p} f_{i} A^{-i} \right) \left( I + \sum_{j=1}^{q} g_{j} A^{-j} \right)^{-1}.
	\nonumber 
	\end{flalign}
	Therefore, it holds that
	\begin{flalign}
	I - \sum_{i=1}^{\infty} h_{i} A^{-i}
	= \left( I - \sum_{i=1}^{p} f_{i} A^{-i} \right) \left( I + \sum_{j=1}^{q} g_{j} A^{-j} \right)^{-1}, \nonumber
	\end{flalign}
    and thus
    \begin{flalign} 
    \widehat{C} 
    &= C \left( I - \sum_{i=1}^{p} f_{i} A^{-i} \right) \left( I + \sum_{j=1}^{q} g_{j} A^{-j} \right)^{-1}, \nonumber
    \end{flalign}
	which completes the proof.
\end{IEEEproof}

We shall now proceed to prove that the system given in \eqref{plant22} is observable (and thus detectable). For simplicity, we will denote (by a slight abuse of notation; cf. \eqref{inverse})
\begin{flalign}
F^{-1} \left( A \right) = \left( I - \sum_{i=1}^{p} f_{i} A^{-i} \right) \left( I + \sum_{j=1}^{q} g_{j} A^{-j} \right)^{-1},
\end{flalign}
for the rest of the paper. Accordingly, \eqref{c33} can be represented as
\begin{flalign}
\widehat{C} 
= C F^{-1} \left( A \right). 
\end{flalign}

\begin{proposition}
	The pair $\left( A, \widehat{C} \right)$ is observable (and thus detectable).
\end{proposition}

\begin{IEEEproof}
	Note that the observation matrix \cite{astrom2010feedback} for $\left( A, \widehat{C} \right)$ is given by
	\begin{flalign}
	&\left[
	\begin{array}{c}
	\widehat{C} \\
	\widehat{C} A \\
	\vdots \\
	\widehat{C} A^{n-1}\\
	\end{array} \right] 
	= \left[
	\begin{array}{c}
	C F^{-1} \left( A \right) \\
	C F^{-1} \left( A \right) A \\
	\vdots \\
	C F^{-1} \left( A \right) A^{n-1}\\
	\end{array} \right] \nonumber \\
	&~~~~ = \left[
	\begin{array}{c}
	C F^{-1} \left( A \right) \\
	C A F^{-1} \left( A \right) \\
	\vdots \\
	C A^{n-1} F^{-1} \left( A \right)\\
	\end{array} \right] 
	= \left[
	\begin{array}{c}
	C \\
	C A \\
	\vdots \\
	C A^{n-1}\\
	\end{array} \right] F^{-1} \left( A \right)
	.\nonumber
	\end{flalign} 
	As such, since  $\left( A, C \right)$
	is observable, i.e., 
	\cite{astrom2010feedback} 
	\begin{flalign}
	\mathrm{rank} \left[
	\begin{array}{c}
		C \\
		C A \\
		\vdots \\
		C A^{n-1}\\
	\end{array} \right] = n,\nonumber
	\end{flalign} 
	it suffices to show that $F^{-1} \left( A \right)$ is invertible in order to show that
	\begin{flalign}
		\mathrm{rank} \left[
		\begin{array}{c}
		\widehat{C} \\
		\widehat{C} A \\
		\vdots \\
		\widehat{C} A^{n-1}\\
		\end{array} \right] = n,\nonumber
	\end{flalign}
	i.e.,
	$\left( A, \widehat{C} \right)$ is also observable.
	To see this, note that it is known from the proof of Proposition~\ref{anderson11} that
	\begin{flalign} \label{matrix}	
	F^{-1} \left( A \right)
	= T \left[
	\begin{array}{cccc}
	\frac{1 - \sum_{i=1}^{p} f_{i} \lambda_{1}^{-i}}{1 + \sum_{j=1}^{q} g_{j} \lambda_{1}^{-j}} & \cdots & 0\\
	\vdots  & \ddots & \vdots\\
	0 & \cdots & \frac{1 - \sum_{i=1}^{p} f_{i} \lambda_{n}^{-i}}{1 + \sum_{j=1}^{q} g_{j} \lambda_{n}^{-j}}\\
	\end{array} \right] T^{-1}.
	\end{flalign}
	Then, since $1 - \sum_{i=1}^{p} f_{i} z^{-i}$ is minimum-phase while the poles of $1 + \sum_{j=1}^{q} g_{j} z^{-j}$ are given by $z_{j} = 0, j = 1, \ldots, q$, it holds for any $\ell = 1, 2, \ldots, n$ that
	\begin{flalign}
	\frac{1 - \sum_{i=1}^{p} f_{i} \lambda_{\ell}^{-i}}{1 + \sum_{j=1}^{q} g_{j} \lambda_{\ell}^{-j}} \neq 0, \nonumber
	\end{flalign}
	due to the fact that $\left| \lambda_{\ell} \right| > 1, \ell = 1, \ldots, n$. Hence, $F^{-1} \left( A \right)$ is invertible, and consequently, $\left( A, \widehat{C} \right)$ is observable (and thus detectable).
\end{IEEEproof}

Meanwhile, the Kalman filter for \eqref{plant22} is given by
\begin{flalign} \label{estimator22}
\left\{ \begin{array}{rcl}
\overline{\mathbf{x}}_{k+1}&=&A \overline{\mathbf{x}}_{k} +\mathbf{u}_k, \\
\overline{\mathbf{y}}_{k}&= & \widehat{C} \overline{\mathbf{x}}_{k}, \\
\mathbf{e}_{k}&=& \widehat{\mathbf{y}}_{k}-\overline{\mathbf{y}}_k, \\
\mathbf{u}_{k}&=& \widehat{K}_{k} \mathbf{e}_{k},
\end{array}
\right.
\end{flalign}
where $\overline{\mathbf{x}}_{k} \in \mathbb{R}^n$, $\overline{\mathbf{y}}_{k} \in \mathbb{R}$, $\mathbf{e}_{k} \in \mathbb{R}$, and $\mathbf{\mathbf{u}}_{k} \in \mathbb{R}^n$. 
\begin{figure}
	\begin{center}
		\vspace{-3mm}
		\includegraphics [width=0.5\textwidth]{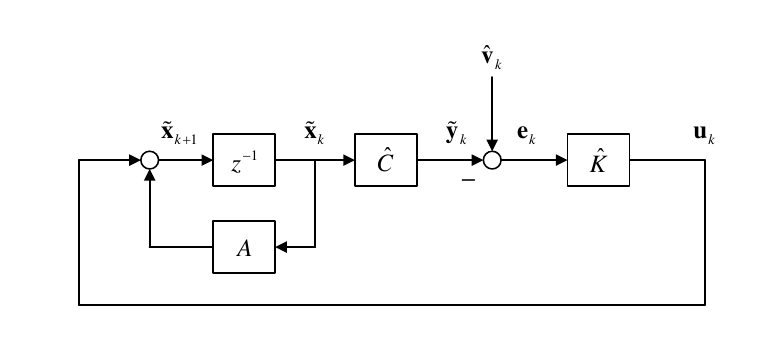}
		\vspace{-6mm}
		\caption{The steady-state integrated Kalman filter for colored noises.}
		\label{kalman3}
	\end{center}
	\vspace{-3mm}
\end{figure}
Furthermore, since $\left( A, \widehat{C} \right)$ is detectable, the Kalman filtering system converges, i.e., the state estimation error $\left\{ \mathbf{x}_{k} - \overline{\mathbf{x}}_{k} \right\}$ is asymptotically stationary. Moreover, in steady state, the optimal state estimation error covariance
\begin{flalign}
P =\lim_{k\to \infty} \mathbb{E} \left[ \left( \mathbf{x}_{k} - \overline{\mathbf{x}}_{k} \right) \left( \mathbf{x}_{k} - \overline{\mathbf{x}}_{k} \right)^{\mathrm{T}} \right] \nonumber
\end{flalign}
attained by the Kalman filter is given by the (non-zero) positive semi-definite solution to the algebraic Riccati equation
\begin{flalign} 
P
= A P A^{\mathrm{T}} - A P \widehat{C}^{\mathrm{T}} \left( \widehat{C} P \widehat{C}^{\mathrm{T}} + \sigma_{\mathbf{v}}^2 \right)^{-1} \widehat{C} P A^{\mathrm{T}}, \nonumber
\end{flalign} 
whereas the steady-state observer gain is given by 
\begin{flalign} \label{gain22}
\widehat{K} = A P \widehat{C}^{\mathrm{T}} \left( \widehat{C} P \widehat{C}^{\mathrm{T}} + \sigma_{\mathbf{v}}^2 \right)^{-1}.
\end{flalign}  

In addition, it holds that (cf. \eqref{bode})
\begin{flalign}
\widehat{C} P \widehat{C}^{\mathrm{T}} + \sigma_{\widehat{\mathbf{v}}}^2 =  \left( \prod_{\ell=1}^{n} \lambda_{\ell}^2 \right) \sigma_{\widehat{\mathbf{v}}}^2,
\end{flalign}
or equivalently,
\begin{flalign} \label{bode2}
\widehat{C} P \widehat{C}^{\mathrm{T}} = \left[ \left( \prod_{\ell=1}^{n} \lambda_{\ell}^2 \right) - 1 \right] \sigma_{\widehat{\mathbf{v}}}^2. 
\end{flalign}

Again, by letting $\widetilde{\mathbf{x}}_{k} = \overline{\mathbf{x}}_{k} - \mathbf{x}_{k} $ and $\widetilde{\mathbf{y}}_{k} = \overline{\mathbf{y}}_{k} - \widehat{\mathbf{z}}_{k} = \overline{\mathbf{y}}_{k} - \widehat{C} \mathbf{x}_{k} $, we may integrate the systems of \eqref{plant22} and \eqref{estimator22} in steady state into an equivalent form:
\begin{flalign} \label{integrate2}
\left\{ \begin{array}{rcl}
\widetilde{\mathbf{x}}_{k+1}&=&A \widetilde{\mathbf{x}}_{k} +\mathbf{u}_k, \\
\widetilde{\mathbf{y}}_{k}&= & \widehat{C} \widetilde{\mathbf{x}}_{k}, \\
\mathbf{e}_{k}&=& - \widetilde{\mathbf{y}}_k + \widehat{\mathbf{v}}_k, \\
\mathbf{u}_{k}&=& \widehat{K} \mathbf{e}_{k},
\end{array}
\right.
\end{flalign}
as depicted in Fig.~\ref{kalman3}. In addition, it may be verified that the closed-loop system given in \eqref{integrate2} and Fig.~\ref{kalman3} is stable \cite{anderson2012optimal, astrom2010feedback}. 

\subsection{Lower Bounds on Feedback Capacity and Feedback Coding}

We now proceed to obtain lower bounds on feedback capacity as well as the corresponding recursive coding schemes to achieve them,
based upon the results and discussions provided in the previous subsection. We first propose a particular way to design $A$ and $C$.

\begin{theorem} \label{t1}
	Suppose that
	\begin{flalign} \label{roots}
	F^{-1} \left( z \right) - \gamma= \frac{1 - \sum_{i=1}^{p} f_{i} z^{-i}}{1 + \sum_{j=1}^{q} g_{j} z^{-j}} - \gamma,
	\end{flalign}
	with a given $\gamma \in \mathbb{R}, \gamma \neq 0$,
	has at least $n \geq 1$ distinct nonminimum-phase zeros. Let
	\begin{flalign} \label{matrixA}
	A = T \left[
	\begin{array}{cccc}
	\lambda_{1} & 0 & \cdots & 0\\
	0 & \lambda_{2} & \cdots & 0\\
	\vdots & \vdots & \ddots & \vdots\\
	0 & 0 & \cdots & \lambda_{n}
	\end{array}
	\right] T^{-1} \in \mathbb{R}^{n \times n},
	\end{flalign}
	where $\lambda_{1}, \lambda_{2}, \ldots, \lambda_{n}, \lambda_{1} \neq \lambda_{2} \neq \cdots \neq \lambda_{n}$, are picked among the nonminimum-phase zeros of \eqref{roots}. Note that conjugate zeros of \eqref{roots}, if there are any, should be picked in pairs in order to render $A$ a real matrix. Note also that herein $T \in \mathbb{R}^{n \times n}$ can be any invertible matrix.
	In addition, choose a $C \in \mathbb{R}^{1 \times n}$ that renders $\left( A, C \right)$ observable, e.g., 
	\begin{flalign} \label{matrixC}
	C = \left[
	\begin{array}{cccc}
	1 & 1 & \cdots & 1
	\end{array}
	\right] T^{-1},
	\end{flalign} 
	Then, in the system of \eqref{integrate2} and Fig.~\ref{kalman3}, 
	it holds for any $\ell = 1, \ldots, n$ that
	\begin{flalign} \label{main} 
	\left( \prod_{\ell=1}^{n} \lambda_{\ell}^2 - 1 \right) \sigma_{\widehat{\mathbf{v}}}^2
	&= \gamma^2 C P C^{\mathrm{T}}   \nonumber \\
	&= \left( \frac{1 - \sum_{i=1}^{p} f_{i} \lambda_{\ell}^{-i}}{1 + \sum_{j=1}^{q} g_{j} \lambda_{\ell}^{-j}} \right)^2 C P C^{\mathrm{T}}.
	\end{flalign}
\end{theorem}

\begin{IEEEproof}
	Suppose that $A$ and $C$ are chosen as in \eqref{matrixA} and \eqref{matrixC}, respectively. Clearly,
	\begin{flalign} \label{bb}
	\frac{1 - \sum_{i=1}^{p} f_{i} \lambda_{1}^{-i}}{1 + \sum_{j=1}^{q} g_{j} \lambda_{1}^{-j}} 
	= \cdots
	= \frac{1 - \sum_{i=1}^{p} f_{i} \lambda_{n}^{-i}}{1 + \sum_{j=1}^{q} g_{j} \lambda_{n}^{-j}} 
	= \gamma,
	\end{flalign} 
	and it then follows from \eqref{matrix} that
	\begin{flalign}
	F^{-1} \left( A \right) 
	&= T \left[
	\begin{array}{cccc}
	\frac{1 - \sum_{i=1}^{p} f_{i} \lambda_{1}^{-i}}{1 + \sum_{j=1}^{q} g_{j} \lambda_{1}^{-j}} & \cdots & 0\\
	\vdots  & \ddots & \vdots\\
	0 & \cdots & \frac{1 - \sum_{i=1}^{p} f_{i} \lambda_{n}^{-i}}{1 + \sum_{j=1}^{q} g_{j} \lambda_{n}^{-j}}\\
	\end{array} \right] T^{-1} \nonumber \\
	&= T \left[
	\begin{array}{cccc}
	\gamma & \cdots & 0\\
	\vdots  & \ddots & \vdots\\
	0 & \cdots & \gamma \\
	\end{array} \right] T^{-1}
	= \gamma I. \nonumber
	\end{flalign}
	As such,
	\begin{flalign} 
	\widehat{C} P \widehat{C}^{\mathrm{T}}
	= C F^{-1} \left( A \right) P F^{-\mathrm{T}} \left( A \right) C^{\mathrm{T}} 
	= \gamma^2 C P C^{\mathrm{T}}. \nonumber
	\end{flalign}
	Meanwhile, note that the pair $\left( A, C \right)$ is observable since the observability matrix is given by
	\begin{flalign}
	\left[ \begin{array}{c}
	C \\
	C A \\
	\vdots \\
	C A^{n-1}\\
	\end{array} \right] = \left[ \begin{array}{cccc}
	1 & 1 & \cdots & 1\\
	\lambda_{1} & \lambda_{2} & \cdots & \lambda_{n}\\
	\vdots & \vdots & \ddots & \vdots\\
	\lambda_{1}^{n-1} & \lambda_{2}^{n-1} & \cdots & \lambda_{n}^{n-1}\\
	\end{array} \right] T^{-1}
	,\nonumber
	\end{flalign} 
	and it can be verified that its rank is $n$, since 
	\begin{flalign}
	\det \left[ \begin{array}{cccc}
	1 & 1 & \cdots & 1\\
	\lambda_{1} & \lambda_{2} & \cdots & \lambda_{n}\\
	\vdots & \vdots & \ddots & \vdots\\
	\lambda_{1}^{n-1} & \lambda_{2}^{n-1} & \cdots & \lambda_{n}^{n-1}\\
	\end{array} \right] = \prod_{0 \leq i < j \leq n} \left( \lambda_{j} - \lambda_{i} \right)
	,\nonumber
	\end{flalign} 
	and $\lambda_{1} \neq \lambda_{2} \neq \cdots \neq \lambda_{n}$. 
	Hence, \eqref{bode2} holds, which can then be rewritten as
	\begin{flalign}
	\gamma^2 C P C^{\mathrm{T}}   
	= \left[ \left( \prod_{\ell=1}^{n} \lambda_{\ell}^2 \right) - 1 \right] \sigma_{\widehat{\mathbf{v}}}^2, \nonumber
	\end{flalign}
	since
	\begin{flalign} 
	\widehat{C} P \widehat{C}^{\mathrm{T}}
	= \gamma^2 C P C^{\mathrm{T}}. \nonumber
	\end{flalign}
	Therefore, \eqref{main} follows by noting also \eqref{bb}.
\end{IEEEproof}

Note in particular the fact that conjugate zeros of \eqref{roots} are included in pairs implicates that
\begin{flalign} \label{real1}
	\prod_{\ell=1}^{n} \left| \lambda_{\ell} \right| = \left| \prod_{\ell=1}^{n} \lambda_{\ell} \right|.
\end{flalign}	
Note also that 
\begin{flalign} \label{max}
n \leq \max \left\{ p, q \right\},
\end{flalign} 
since \eqref{roots} has at most $\max \left\{ p, q \right\}$ zeros.

\begin{figure}
	\begin{center}
		\vspace{-3mm}
		\includegraphics [width=0.5\textwidth]{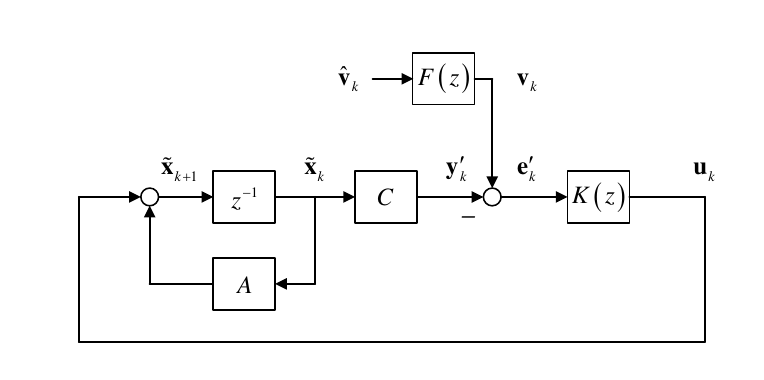}
		\vspace{-6mm}
		\caption{The steady-state integrated Kalman filter for colored noises: Equivalent form.}
		\label{kalman4}
	\end{center}
	\vspace{-3mm}
\end{figure}

Meanwhile, we may obtain an equivalent form of the system given in \eqref{integrate2} and Fig.~\ref{kalman3}.

\begin{theorem} \label{observation3}
	The system in Fig.~\ref{kalman3} is equivalent to that in Fig.~\ref{kalman4}, where $K \left( z \right)$ is dynamic and is given as
	\begin{flalign}
	K \left( z \right)
	= F^{-1} \left( z \right) \widehat{K} 
	=  \frac{1 - \sum_{i=1}^{p} f_{i} z^{-i}}{1 + \sum_{j=1}^{q} g_{j} z^{-j}} \widehat{K}. 
	\end{flalign}
	Herein, $\widehat{K}$ is given by \eqref{gain22}.
	More specifically, the system in Fig.~\ref{kalman4} is given by 
	\begin{flalign} \label{integrate6}
	\left\{ \begin{array}{rcl}
	\widetilde{\mathbf{x}}_{k+1}&=&A \widetilde{\mathbf{x}}_{k} +\mathbf{u}_k, \\
	\mathbf{y}'_{k}&= & C \widetilde{\mathbf{x}}_{k}, \\
	\mathbf{e}'_{k}&=& - \mathbf{y}'_k + \mathbf{v}_k, \\
	\mathbf{u}_{k}&=& \widehat{K} \left( \mathbf{e}'_{k} - \sum_{i=1}^{p} f_{i} \mathbf{e}'_{k-i} \right) - \sum_{j=1}^{q} g_{j} \mathbf{u}_{k-j},
	\end{array}
	\right.
	\end{flalign}
	which is stable as a closed-loop system.
\end{theorem}

\begin{IEEEproof}
	Note first that the system in Fig.~\ref{kalman3} is equivalent to the one in Fig.~\ref{kalman5} since $\widehat{K}  = F \left( z \right) K \left( z \right)$.
	\begin{figure}
		\begin{center}
			\vspace{-3mm}
			\includegraphics [width=0.5\textwidth]{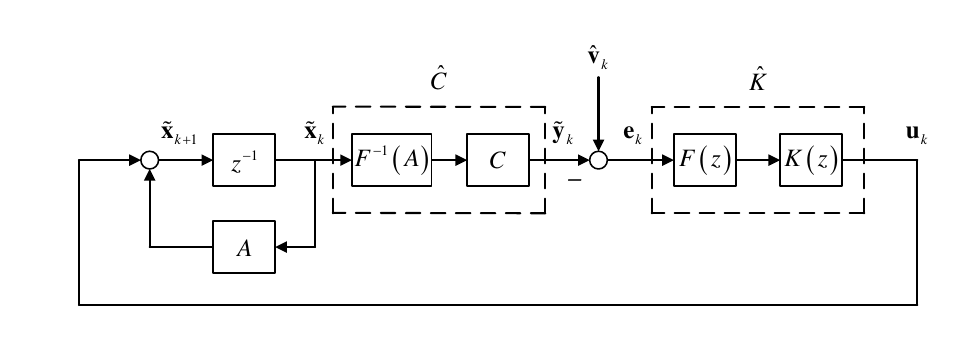}
			\vspace{-6mm}
			\caption{The steady-state integrated Kalman filter for colored noises: Equivalent form~2.}
			\label{kalman5}
		\end{center}
		\vspace{-3mm}
	\end{figure}
	Then, since 
	\begin{flalign}
	\widetilde{\mathbf{x}}_{k} = \overline{\mathbf{x}}_{k} - \mathbf{x}_{k} = A \left( \overline{\mathbf{x}}_{k-1} - \mathbf{x}_{k-1}\right) = A \widetilde{\mathbf{x}}_{k-1}, \nonumber
	\end{flalign}
	we have
	\begin{flalign}
	\widetilde{\mathbf{y}}_{k} 
	&= \widehat{C} \widetilde{\mathbf{x}}_{k} =  C F^{-1} \left( A \right) \widetilde{\mathbf{x}}_{k} \nonumber \\
	&= C \left( I - \sum_{i=1}^{p} f_{i} A^{-i} \right) \left( I + \sum_{j=1}^{q} g_{j} A^{-j} \right)^{-1} \widetilde{\mathbf{x}}_{k} \nonumber \\
	&= C \left( I - \sum_{i=1}^{\infty} h_{i} A^{-i} \right) \widetilde{\mathbf{x}}_{k}
	= C \left( 1 - \sum_{i=1}^{\infty} h_{i} z^{-i} \right) \widetilde{\mathbf{x}}_{k} \nonumber \\
	&= C \left( \frac{1 - \sum_{i=1}^{p} f_{i} z^{-i}}{1 + \sum_{j=1}^{q} g_{j} z^{-j}} \right) \widetilde{\mathbf{x}}_{k} \nonumber \\
	&= C F^{-1} \left( z \right) \widetilde{\mathbf{x}}_{k} = F^{-1} \left( z \right) C  \widetilde{\mathbf{x}}_{k}. \nonumber
	\end{flalign}
	(Note that this ``static-dynamic equivalence'' transformation, where $C F^{-1} \left( A \right)$ is static and $F^{-1} \left( z \right) C $ is dynamic, is a critical step.)
	Consequently, the system of Fig.~\ref{kalman5} is equivalent to that of Fig.~\ref{kalman6}.
	\begin{figure}
		\begin{center}
			\vspace{-3mm}
			\includegraphics [width=0.5\textwidth]{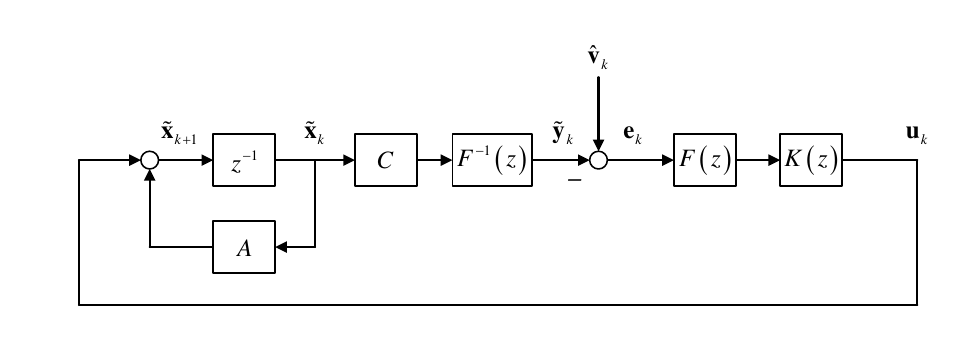}
			\vspace{-6mm}
			\caption{The steady-state integrated Kalman filter for colored noises: Equivalent form~3.}
			\label{kalman6}
		\end{center}
		\vspace{-3mm}
	\end{figure}
	Moreover, since all the sub-systems are linear, the system of Fig.~\ref{kalman6} is equivalent to that of Fig.~\ref{kalman7}, which in turn equals to the one of Fig.~\ref{kalman4}; note that herein $F \left( z \right)$ is stable and minimum-phase, and thus there will be no issues caused by cancellations of unstable poles and nonminimum-phase zeros.
	\begin{figure}
		\begin{center}
			\vspace{-3mm}
			\includegraphics [width=0.5\textwidth]{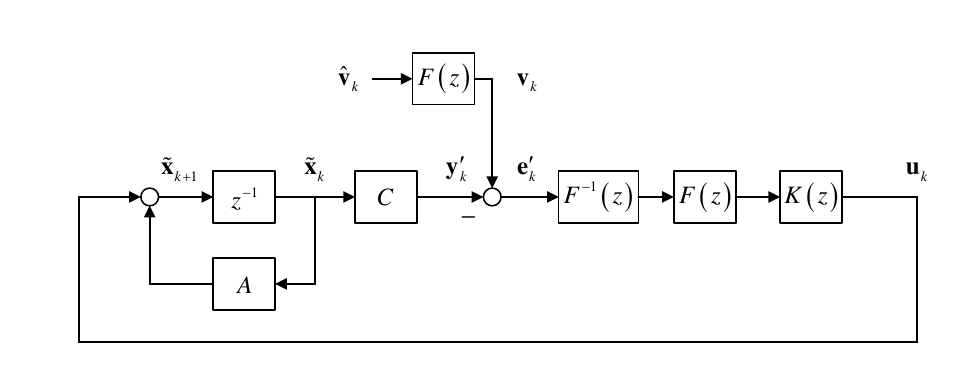}
			\vspace{-6mm}
			\caption{The steady-state integrated Kalman filter for colored noises: Equivalent form~4.}
			\label{kalman7}
		\end{center}
		\vspace{-3mm}
	\end{figure}
	Meanwhile, the closed-loop stability of the system given in \eqref{integrate6} and Fig.~\ref{kalman4} is the same as that of the system given by \eqref{integrate2} and Fig.~\ref{kalman3}, since they are essentially the same feedback system.
\end{IEEEproof}


\begin{figure}
	\begin{center}
		\vspace{-3mm}
		\includegraphics [width=0.5\textwidth]{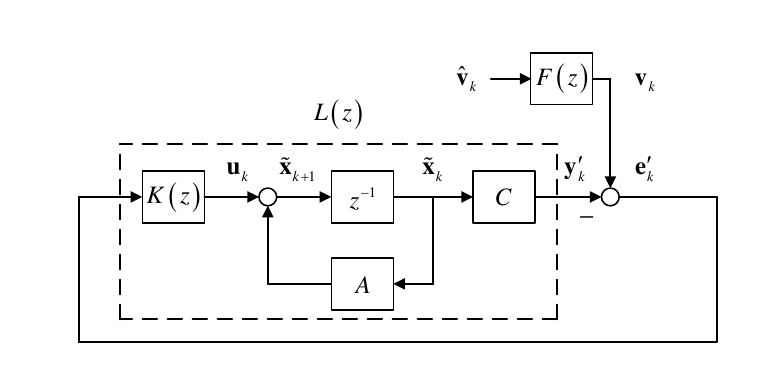}
		\vspace{-6mm}
		\caption{The steady-state integrated Kalman filter for colored noises: Equivalent form~5.}
		\label{kalman10}
	\end{center}
	\vspace{-3mm}
\end{figure} 

Note in particular that in the system of \eqref{integrate6} and Fig.~\ref{kalman4}, it holds that
\begin{flalign} \label{constraint}
\mathbb{E} \left[ \left( \mathbf{y}'_{k} \right)^2 \right] 
&= \mathbb{E} \left[ \left( C \widetilde{\mathbf{x}}_{k} \right)^2 \right] 
= \mathbb{E} \left[ \left( C \overline{\mathbf{x}}_{k} - C \mathbf{x}_{k} \right)^2 \right] \nonumber \\
&= C \left\{ \mathbb{E} \left[ \left(  \overline{\mathbf{x}}_{k} - \mathbf{x}_{k} \right) \left(  \overline{\mathbf{x}}_{k} - \mathbf{x}_{k} \right)^{\mathrm{T}} \right] \right\} C^{\mathrm{T}}
= C P C^{\mathrm{T}},
\end{flalign} 
since the system is LTI and $\left\{ \mathbf{x}_k -\overline{\mathbf{x}}_k \right\}$ is stationary.

We next provide a key observation that enables relating the Kalman filtering system to the feedback capacity.

\begin{observation} \label{observation1}
	In the system of \eqref{integrate6} and Fig.~\ref{kalman4}, or equivalently, in the system of Fig.~\ref{kalman10}, we may view 
	\begin{flalign} \label{notations}
	\mathbf{e}'_{k} = - \mathbf{y}'_{k} + \mathbf{v}_{k}
	\end{flalign}
	as a feedback channel \cite{kim2010feedback, ardestanizadeh2012control} with ARMA noise $\left\{ \mathbf{v}_{k} \right\}$, whereas $\left\{ - \mathbf{y}'_{k} \right\}$ is the channel input while $\left\{ \mathbf{e}'_{k} \right\}$ is the channel output. Moreover, as illustrated in Fig.~\ref{kalman10}, 
	\begin{flalign} \label{feedbackcoding}
	L \left( z \right) = C \left( z I - A \right)^{-1} K \left( z \right)
	\end{flalign}
	may be viewed as the feedback coding scheme; cf. \cite{ardestanizadeh2012control}, or alternatively,
	\cite{kim2010feedback} with 
	\begin{flalign} \label{feedbackcoding2}
	B \left( z \right)
	= \frac{- L \left( z \right)}{ 1 + L \left( z \right)} = \frac{- C \left( z I - A \right)^{-1} K \left( z \right)}{1 + C \left( z I - A \right)^{-1} K \left( z \right)}.
	\end{flalign}
	Meanwhile, with the notations in \eqref{notations}, the feedback capacity is given by (cf. the definition in \eqref{fcdef})
	\begin{flalign} \label{notations2}
	C_{\text{f}} = \sup_{ \left\{ - \mathbf{y}'_{k} \right\} } \left[ h_{\infty} \left( \mathbf{e}' \right) -  h_{\infty} \left(  \mathbf{v} \right) \right],
	\end{flalign}
	where the supremum is taken over all stationary channel input processes $\left\{ - \mathbf{y}'_{k} \right\}$ of the form 
	\begin{flalign} \label{linear}
	- \mathbf{y}'_{k} = \sum_{i=1}^{\infty} b_{i} \mathbf{v}_{k-i},~b_{i} \in \mathbb{R}, 
	\end{flalign} 
	while satisfying
	\begin{flalign} \label{power}
	\mathbb{E} \left[ \left( - \mathbf{y}'_{k} \right)^2 \right] = \mathbb{E} \left[ \left( \mathbf{y}'_{k} \right)^2 \right] \leq \overline{P}.
	\end{flalign} 
	As such, if $A$ and $C$ are designed specifically as in Theorem~\ref{t1}, then \eqref{feedbackcoding} provides a class of sub-optimal feedback coding scheme as long as \eqref{power} is satisfied, and the corresponding 
	\begin{flalign}
		h_{\infty} \left( \mathbf{e}' \right) -  h_{\infty} \left(  \mathbf{v} \right)
	\end{flalign}
	is thus a lower bound of \eqref{notations2}. 
	
	\begin{IEEEproof}	
	Note that herein we have used the fact that $\left\{ - \mathbf{y}'_{k} \right\}$ is stationary and $- \mathbf{y}'_{k}$ is a linear combination of the past $\left\{ \mathbf{v}_{k} \right\}$ up to time $k-1$, i.e., \eqref{linear} holds. 
	To see this, note first that, according to
	Theorem~\ref{observation3}, the system in \eqref{integrate6} and Fig.~\ref{kalman4} is stable (and LTI). In particular,
	the transfer function from $\left\{ \mathbf{v}_{k} \right\}$ to $\left\{ - \mathbf{y}'_{k} \right\}$, given as
	\begin{flalign} \label{transfer}
	\frac{- C \left( z I - A \right)^{-1} K \left( z \right)}{1 + C \left( z I - A \right)^{-1} K \left( z \right)} 
	\end{flalign}
	is stable (and LTI).
	Then, since $\left\{ \mathbf{v}_{k} \right\}$ is stationary, $\left\{ - \mathbf{y}'_{k} \right\}$ is also stationary.	
	On the other hand, it can be verified that \eqref{transfer}
	is strictly causal, since 
	\begin{flalign}
	K \left( z \right) 
	=  F^{-1} \left( z \right)  \widehat{K} 
	=  \frac{1 - \sum_{i=1}^{p} f_{i} z^{-i}}{1 + \sum_{j=1}^{q} g_{j} z^{-j}} \widehat{K} \nonumber
	\end{flalign}
	is causal. As a result, \eqref{linear} holds.
	\end{IEEEproof}
	
\end{observation}


Based upon Observation~\ref{observation1}, the following lower bound of feedback capacity
can be obtained.


\begin{theorem} \label{fc4}
	Given an $n \geq 1$, suppose that
	\begin{flalign} \label{solution}
	\left\{ \begin{array}{rcl}
	1 - \sum_{i=1}^{p} f_{i} \lambda_{1}^{-i} & = & \zeta \left( 1 + \sum_{j=1}^{q} g_{j} \lambda_{1}^{-j} \right),\\
	& \vdots & \\
	1 - \sum_{i=1}^{p} f_{i} \lambda_{n}^{-i} & = & \zeta \left( 1 + \sum_{j=1}^{q} g_{j} \lambda_{n}^{-j} \right),\\
	\zeta^2 \frac{\overline{P}}{\sigma_{\widehat{\mathbf{v}}}^2} & = &  \prod_{\ell=1}^{n} \lambda_{\ell}^2  - 1,
	\end{array} \right. 
	\end{flalign}
	has at least one solution $\left[ \lambda_{1}, \ldots, \lambda_{n} \right]^{\mathrm{T}}$
	that satisfies $\left| \lambda_{\ell} \right| > 1, \ell = 1, \ldots, n$ while $\lambda_{1} \neq \lambda_{2} \neq \cdots \neq \lambda_{n}$ and $\prod_{\ell=1}^{n} \lambda_{\ell} \in \mathbb{R}$; note that herein $\zeta \in \mathbb{R}, \zeta \neq 0$, is an auxiliary parameter.
	Among all such solutions, denote the one that provides the largest $\left| \prod_{\ell=1}^{n} \lambda_{\ell} \right|$ as $\left[ \overline{\lambda}_{1}, \ldots, \overline{\lambda}_{n} \right]^{\mathrm{T}}$.
	Then, a lower bound of the feedback capacity with ARMA noise \eqref{ARMA} and power constraint $\overline{P}$ is given by 
	\begin{flalign} \label{solution2}
	\log \left| \prod_{\ell=1}^{n} \overline{\lambda}_{\ell} \right|.
	\end{flalign}	
\end{theorem}

\begin{IEEEproof}
	Note first it is supposed that 
	\eqref{solution} has at least one solution $\left[ \lambda_{1}, \ldots, \lambda_{n} \right]^{\mathrm{T}}$
	that satisfies $\left| \lambda_{\ell} \right| > 1, \ell = 1, \ldots, n$ while $\lambda_{1} \neq \lambda_{2} \neq \cdots \neq \lambda_{n}$, which implicates that  $\lambda_{1}, \lambda_{2}, \ldots, \lambda_{n}$ are in fact distinctive nonminimum-phase zeros of
	\begin{flalign} 
	F^{-1} \left( z \right) - \gamma
	= \frac{1 - \sum_{i=1}^{p} f_{i} z^{-i}}{1 + \sum_{j=1}^{q} g_{j} z^{-j}} - \gamma. \nonumber
	\end{flalign}
	In addition, the fact that $\prod_{\ell=1}^{n} \lambda_{\ell} \in \mathbb{R}$ implicates that  conjugate zeros, if there are any, are included herein in pairs. We may then let
	\begin{flalign}
	A = T \left[
	\begin{array}{cccc}
	\lambda_{1} & 0 & \cdots & 0\\
	0 & \lambda_{2} & \cdots & 0\\
	\vdots & \vdots & \ddots & \vdots\\
	0 & 0 & \cdots & \lambda_{n}
	\end{array}
	\right] T^{-1} \in \mathbb{R}^{n \times n}, \nonumber
	\end{flalign}
	where $T \in \mathbb{R}^{n \times n}$ can be any invertible matrix,
	and choose a $C \in \mathbb{R}^{1 \times n}$ that renders $\left( A, C \right)$ observable (see the proof of Theorem~\ref{t1}), e.g., 
	\begin{flalign}
	C = \left[
	\begin{array}{cccc}
	1 & 1 & \cdots & 1
	\end{array}
	\right] T^{-1}. \nonumber
	\end{flalign} 
	Hence, in the system of \eqref{integrate2} and Fig.~\ref{kalman3},
	it holds for any $\ell = 1, \ldots, n$ that (see Theorem~\ref{t1})
	\begin{flalign} 
	\left( \prod_{\ell=1}^{n} \lambda_{\ell}^2 - 1 \right) \sigma_{\widehat{\mathbf{v}}}^2
	&= \zeta^2 C P C^{\mathrm{T}}   \nonumber \\
	&= \left( \frac{1 - \sum_{i=1}^{p} f_{i} \lambda_{\ell}^{-i}}{1 + \sum_{j=1}^{q} g_{j} \lambda_{\ell}^{-j}} \right)^2 C P C^{\mathrm{T}}. \nonumber
	\end{flalign}
	On the other hand, it is known from \eqref{solution} that
	\begin{flalign}
		\zeta^2 \frac{\overline{P}}{\sigma_{\widehat{\mathbf{v}}}^2}  =   \prod_{\ell=1}^{n} \lambda_{\ell}^2  - 1, \nonumber
	\end{flalign}
	or equivalently, 
	\begin{flalign}
	\left( \prod_{\ell=1}^{n} \lambda_{\ell}^2 - 1 \right) \sigma_{\widehat{\mathbf{v}}}^2
	= \zeta^2 \overline{P}. \nonumber
	\end{flalign}
	Thus,
	\begin{flalign}
	C P C^{\mathrm{T}} = \overline{P} = \left( \prod_{\ell=1}^{n} \lambda_{\ell}^2 - 1 \right) \frac{\sigma_{\widehat{\mathbf{v}}}^2}{\zeta^2}, \nonumber
	\end{flalign}
	which means that the power constraint of \eqref{power} is satisfied in the system of \eqref{integrate6} and Fig.~\ref{kalman4}, since it is known from \eqref{constraint} that 
	\begin{flalign} 
	\mathbb{E} \left[ \left( \mathbf{y}'_{k} \right)^2 \right] 
	= C P C^{\mathrm{T}}. \nonumber
	\end{flalign} 
	
	As such, according to Observation~\ref{observation1}, the corresponding \eqref{feedbackcoding} provides a class of sub-optimal feedback coding, whereas the corresponding 	
	\begin{flalign}
	h_{\infty} \left( \mathbf{e}' \right) -  h_{\infty} \left(  \mathbf{v} \right) \nonumber
	\end{flalign} 
	provides a lower bound of the feedback capacity.
	In addition, since the transfer function from $\left\{ \mathbf{v}_{k} \right\}$ to $\left\{ \mathbf{e}'_{k} \right\}$ is given by
	\begin{flalign}
	\frac{1}{1 + C \left( z I - A \right)^{-1} K \left( z \right)}, \nonumber
	\end{flalign}
	it holds that (cf. discussions in \cite{Eli:04, ardestanizadeh2012control}),
	\begin{flalign}
	&h_{\infty} \left( \mathbf{e}' \right) - h_{\infty} \left( \mathbf{v} \right) \nonumber \\
	&~~~~ = \frac{1}{2 \pi} \int_{- \pi}^{\pi} \log \left| \frac{1}{1 + C \left( \mathrm{e}^{\mathrm{j} \omega} I - A \right)^{-1} K \left( \mathrm{e}^{\mathrm{j} \omega} \right)} \right| \mathrm{d} \omega \nonumber \\
	&~~~~ = \frac{1}{2 \pi} \int_{- \pi}^{\pi} \log \left| \frac{1}{1 + C \left( \mathrm{e}^{\mathrm{j} \omega} I - A \right)^{-1} F^{-1} \left( \mathrm{e}^{\mathrm{j} \omega} \right) \widehat{K}} \right| \mathrm{d} \omega 
	\nonumber \\
	&~~~~ = \sum_{\ell=1}^{n} \log \left| \lambda_{\ell} \right| = \log \prod_{\ell=1}^{n} \left| \lambda_{\ell} \right| = \log \left| \prod_{\ell=1}^{n} \lambda_{\ell} \right|
	, \nonumber
	\end{flalign}
	where the first equality may be referred to \cite{Pap:02, fang2017towards}, the third equality follows as a result of the Bode integral or Jensen's formula \cite{seron2012fundamental, fang2017towards}, and the last equality since $A$ is a real matrix (see \eqref{real1}). Note that herein we have used the fact that $F^{-1} \left( z \right)$ is stable and minimum-phase and 
	\begin{flalign}
	\frac{1}{1 + C \left( z I - A \right)^{-1} K \left( z \right)} \nonumber
	\end{flalign}
	is stable. We have also utilized the fact that  $\left( A, C \right)$ is controllable, and thus the set of unstable poles of 
	\begin{flalign}
	C \left( z I - A \right)^{-1} K \left( z \right)
	\nonumber
	\end{flalign}
	is exactly the same as the set of eigenvalues of $A$ with magnitudes greater than or equal to $1$ \cite{astrom2010feedback}; see also, e.g., discussions in \cite{FangACC18}.
	As such, the lower bound on feedback capacity is equivalently given by 
	\begin{flalign}
	\log \left| \prod_{\ell=1}^{n} \lambda_{\ell} \right|. \nonumber
	\end{flalign}
	Meanwhile, we may pick the solution
	$\left[ \overline{\lambda}_{1}, \ldots, \overline{\lambda}_{n} \right]^{\mathrm{T}}$ that denotes the one that provides the largest $\left| \prod_{\ell=1}^{n} \lambda_{\ell} \right|$, that is, the largest $\log \left| \prod_{\ell=1}^{n} \lambda_{\ell} \right|$, among all the solutions of \eqref{solution} 
	that satisfy $\left| \lambda_{\ell} \right| > 1, \ell = 1, \ldots, n$ while $\lambda_{1} \neq \lambda_{2} \neq \cdots \neq \lambda_{n}$ and $\prod_{\ell=1}^{n} \lambda_{\ell} \in \mathbb{R}$. Accordingly, the lower bound on feedback capacity is given by
	\begin{flalign}
	\log \left| \prod_{\ell=1}^{n} \overline{\lambda}_{\ell} \right|. \nonumber
	\end{flalign}
	This completes the proof.
\end{IEEEproof}

Equivalently, the lower bound can be rewritten, in terms of the ``signal-to-noise" ratio $\overline{P} / \sigma_{\widehat{\mathbf{v}}}^2$, as
\begin{flalign} \label{equal}
\frac{1}{2} \log \left[ 1 + \left( \frac{1 - \sum_{i=1}^{p} f_{i} \overline{\lambda}_{\ell}^{-i}}{1 + \sum_{j=1}^{q} g_{j} \overline{\lambda}_{\ell}^{-j}} \right)^2 \frac{\overline{P}}{\sigma_{\widehat{\mathbf{v}}}^2} \right],~\forall \ell \in 1, \ldots, n,
\end{flalign}
since
it holds for any $\ell = 1, \ldots, n$ that (see the proof of Theorem~\ref{fc4})
\begin{flalign} 
\left( \prod_{\ell=1}^{n} \overline{\lambda}_{\ell}^2 - 1 \right) \sigma_{\widehat{\mathbf{v}}}^2
= \zeta^2 \overline{P}  
= \left( \frac{1 - \sum_{i=1}^{p} f_{i} \overline{\lambda}_{\ell}^{-i}}{1 + \sum_{j=1}^{q} g_{j} \overline{\lambda}_{\ell}^{-j}} \right)^2 \overline{P}, \nonumber
\end{flalign}
whereas
\begin{flalign}
\log \left| \prod_{\ell=1}^{n} \overline{\lambda}_{\ell} \right|
= \log \prod_{\ell=1}^{n} \left| \overline{\lambda}_{\ell} \right|
= \frac{1}{2} \log \prod_{\ell=1}^{n} \left| \overline{\lambda}_{\ell} \right|^2
= \frac{1}{2} \log \prod_{\ell=1}^{n}  \overline{\lambda}_{\ell}^2.
\end{flalign}

In fact, \eqref{max} indicates that it is unnecessary to employ any $n > \max \left\{ p, q \right\}$ in Theorem~\ref{fc4}. Accordingly, we may obtain the following combined lower bound based on Theorem~\ref{fc4}.

\begin{corollary}
	A combined lower bound of the feedback capacity with ARMA noise \eqref{ARMA} and power constraint $\overline{P}$ is given by
	\begin{flalign}
	\max_{1 \leq n \leq \max \left\{ p, q \right\}} \log \left| \prod_{\ell=1}^{n} \overline{\lambda}_{\ell} \right|.
	\end{flalign}	
	where $\overline{\lambda}_{\ell}, \ell = 1, \ldots, n$, are given as in Theorem~\ref{fc4}.
\end{corollary}

We now show that the combined lower bound always exists by simply verifying the lower bound in Theorem~\ref{fc4} always exists for $n = 1$.

\begin{corollary} \label{n1}
	Consider the special case of $n=1$.
	Suppose that
	\begin{flalign} \label{n11}
	\left( \frac{1 - \sum_{i=1}^{p} f_{i} \lambda^{-i}}{1 + \sum_{j=1}^{q} g_{j} \lambda^{-j}} \right)^2 \frac{\overline{P}}{\sigma_{\widehat{\mathbf{v}}}^2}  =   \lambda^2  - 1 
	\end{flalign}
	has at least one root $\lambda$
	that satisfies $\left| \lambda \right| > 1, \lambda \in \mathbb{R}$. 
	Among all such roots, denote the one with the absolute value as $\overline{\lambda}$.
	Then, a lower bound of the feedback capacity with ARMA noise \eqref{ARMA} and power constraint $\overline{P}$ is given by 
	\begin{flalign} \label{n12}
	\log \left| \overline{\lambda} \right|,
	\end{flalign}	
	which is equal to
	\begin{flalign} \label{n13}
	\frac{1}{2} \log \left[ 1 + \left( \frac{1 - \sum_{i=1}^{p} f_{i} \overline{\lambda}^{-i}}{1 + \sum_{j=1}^{q} g_{j} \overline{\lambda}^{-j}} \right)^2 \frac{\overline{P}}{\sigma_{\widehat{\mathbf{v}}}^2} \right].
	\end{flalign}
	In addition, the lower bound always exists, i.e., \eqref{n11} does have at least one real root $\lambda$ that satisfies $\left| \lambda \right| > 1$.
\end{corollary}

\begin{IEEEproof}
	It is clear that when $n = 1$, \eqref{solution} and \eqref{solution2} reduce to \eqref{n11} and \eqref{n12}, respectively, whereas \eqref{equal} becomes \eqref{n13}. To show that \eqref{n11} does have at least one real root that satisfies $\left| \lambda \right| > 1$, note first that when $\lambda = 1$, 
	\begin{flalign} 
	&\left( \frac{1 - \sum_{i=1}^{p} f_{i} \lambda^{-i}}{1 + \sum_{j=1}^{q} g_{j} \lambda^{-j}} \right)^2 \frac{\overline{P}}{\sigma_{\widehat{\mathbf{v}}}^2}  -  \lambda^2  + 1 \nonumber \\
	&~~~~ =  
	\left( \frac{1 - \sum_{i=1}^{p} f_{i}}{1 + \sum_{j=1}^{q} g_{j} } \right)^2 \frac{\overline{P}}{\sigma_{\widehat{\mathbf{v}}}^2}  -  1  + 1 > 0, \nonumber
	\end{flalign}
	whereas 
	\begin{flalign} 
	&\lim_{\lambda \to \infty} \left[ \left( \frac{1 - \sum_{i=1}^{p} f_{i} \lambda^{-i}}{1 + \sum_{j=1}^{q} g_{j} \lambda^{-j}} \right)^2 \frac{\overline{P}}{\sigma_{\widehat{\mathbf{v}}}^2}  -  \lambda^2  + 1 \right] \nonumber \\
	&~~~~ =  \frac{\overline{P}}{\sigma_{\widehat{\mathbf{v}}}^2}  -  \infty  + 1 < 0. \nonumber
	\end{flalign}
	Herein, we have used the fact that 
	\begin{flalign} 
	\frac{1 - \sum_{i=1}^{p} f_{i} z^{-i}}{1 + \sum_{j=1}^{q} g_{j} z^{-j}} \nonumber
	\end{flalign}
	is minimum-phase, and hence for $z = 1$,
	\begin{flalign} 
	\frac{1 - \sum_{i=1}^{p} f_{i} z^{-i}}{1 + \sum_{j=1}^{q} g_{j} z^{-j}} \neq 0,~
	\left( \frac{1 - \sum_{i=1}^{p} f_{i} z^{-i}}{1 + \sum_{j=1}^{q} g_{j} z^{-j}} \right)^2 > 0. \nonumber
	\end{flalign}
	In addition, since 
	\begin{flalign} 
	\frac{1 - \sum_{i=1}^{p} f_{i} z^{-i}}{1 + \sum_{j=1}^{q} g_{j} z^{-j}} \nonumber
	\end{flalign}
	is stable, indicating that 
	\begin{flalign} 
	\left( \frac{1 - \sum_{i=1}^{p} f_{i} \lambda^{-i}}{1 + \sum_{j=1}^{q} g_{j} \lambda^{-j}} \right)^2 \frac{\overline{P}}{\sigma_{\widehat{\mathbf{v}}}^2}  -  \lambda^2  + 1 \nonumber
	\end{flalign}
	is continuous on $\left( 1, \infty \right)$, it follows that \eqref{n11} has at least one root within $\left( 1, \infty \right)$, i.e., a real root with $\left| \lambda \right| > 1$.
\end{IEEEproof}

In particular, for AR noises, i.e., when $g_{j}=0,~j= 1, \ldots, q$, the lower bound in Corollary~\ref{n1} reduces to
the lower bound of \cite{Eli:04} (see Section~V.B therein), which also discussed the relation to the previous lower bound obtained in \cite{butman1976linear}. 

We next consider the special case of first-order ARMA noises with $p = q = 1$. In this case, since $\max \left\{ p, q \right\} = 1$, it suffices to consider only $n = 1$, and accordingly, Example~\ref{casethree} follows directly from Corollary~\ref{n1}. 

\begin{example} \label{casethree}
	In particular, when $p = q = 1$, a lower bound of the feedback capacity with the first-order ARMA noise
	\begin{flalign}
	\mathbf{v}_{k}  
	=   f_{1} \mathbf{v}_{k-1} + \widehat{\mathbf{v}}_k + g_{1} \widehat{\mathbf{v}}_{k-1}, \nonumber
	\end{flalign}
	assuming that
	\begin{flalign}
		\frac{1 + g_{1} z^{-1}}{1 - f_{1} z^{-1}} \nonumber
	\end{flalign}
	is stable and minimum-phase,
	is given by 
	\begin{flalign}
	\log \left| \overline{\lambda} \right|,
	\end{flalign}
	where $\overline{\lambda}$ denotes the real root of 
	\begin{flalign} \label{example}
	\left( \frac{1 - f_{1} \lambda^{-1}}{1 + g_{1} \lambda^{-1}} \right)^2 \frac{\overline{P}}{\sigma_{\widehat{\mathbf{v}}}^2}  = \lambda^2 - 1 
	\end{flalign}
	with the largest absolute value; note that it has been proved in Corollary~\ref{n1} that \eqref{example} has at least one real root with absolute value greater than $1$. Moreover, the lower bound may be rewritten as
	\begin{flalign}
	\log \overline{\lambda},
	\end{flalign}
	where $\overline{\lambda}$ denotes the largest positive real root to be chosen among the roots of 
	\begin{flalign}
	\left( \frac{1 - f_{1} \lambda^{-1}}{1 + g_{1} \lambda^{-1}} \right)^2 \frac{\overline{P}}{\sigma_{\widehat{\mathbf{v}}}^2}  = \lambda^2 - 1,
	\end{flalign}
	as well as
	\begin{flalign}
	\left( \frac{1 + f_{1} \lambda^{-1}}{1 - g_{1} \lambda^{-1}} \right)^2 \frac{\overline{P}}{\sigma_{\widehat{\mathbf{v}}}^2}  = \lambda^2 - 1.
	\end{flalign}
\end{example}

Example~\ref{casethree} is consistent with the feedback capacity of such channels derived in \cite{kim2010feedback} (see also \cite{liu2018feedback}). In fact, it has been shown in \cite{kim2010feedback} (see Theorem~5.3 therein) that the feedback capacity, rather than its lower bound, is given by the unique (therefore, the largest) positive real root of 
\begin{flalign} \label{case1}
	\left( \frac{1 - f_{1} \lambda^{-1}}{1 + g_{1} \lambda^{-1}} \right)^2 \frac{\overline{P}}{\sigma_{\widehat{\mathbf{v}}}^2}  = \lambda^2 - 1,
\end{flalign}
when $f_{1} + g_{1} \leq 0$, while when $f_{1} + g_{1} > 0$, the feedback capacity is given by the only positive real root of 
\begin{flalign} \label{case2}
	\left( \frac{1 + f_{1} \lambda^{-1}}{1 - g_{1} \lambda^{-1}} \right)^2 \frac{\overline{P}}{\sigma_{\widehat{\mathbf{v}}}^2}  = \lambda^2 - 1.
\end{flalign}
Note that it may be verified that when $f_{1} + g_{1} \leq 0$, the positive real root of \eqref{case1} is larger than or equal to that of \eqref{case2}, and vice versa. That is to say, the lower bound in Example~\ref{casethree} is indeed tight. (It is also worth mentioning that before \cite{kim2010feedback}, lower bounds on feedback capacity for the first-order AR noises and the first-order ARMA noises have been respectively obtained in \cite{butman1969general, butman1976linear} and \cite{yang2007feedback}. Meanwhile, see \cite{kim2006feedback, ihara2019feedback} for further discussions on the feedback capacity for the first-order MA noises.)

We next consider the special case of second-order MA noises.

\begin{example} \label{case}
	By letting $p = 0$ and $q = 2$, we may now obtain lower bounds on the feedback capacity with the second-order MA noise
	\begin{flalign} \label{machannel}
	\mathbf{v}_{k}  
	=    \widehat{\mathbf{v}}_k + g_2 \widehat{\mathbf{v}}_{k-2}, 
	\end{flalign}
	assuming that $1 + g_2 z^{-2}$ is minimum-phase. In addition, since $\max \left\{ p, q \right\} = 2$, it suffices to consider $n = 1$ and $n = 2$.
	\begin{itemize}
		\item Case 1: $n = 1$. In this case, Corollary~\ref{n1} will give the following lower bound:
		\begin{flalign}
		\log \left| \overline{\lambda} \right|,
		\end{flalign}
		where $\overline{\lambda}$ denotes the real root of
		\begin{flalign} 
		\left( \frac{1}{1 + g_2 \lambda^{-2}} \right)^2 \frac{\overline{P}}{\sigma_{\widehat{\mathbf{v}}}^2}  =   \lambda^2  - 1
		\end{flalign}
		with the largest absolute value.
		According to \cite{kim2006feedback} (see Section~IV therein), this lower bound is not tight.
		\item Case 2: $n = 2$. In this case, 
		\eqref{solution} reduces to 
		\begin{flalign}  
		\left\{ \begin{array}{rcl}
		1 & = &  \zeta \left( 1 + g_2 \lambda_{1}^{-2} \right),\\
		1 & = &  \zeta \left( 1 + g_2 \lambda_{2}^{-2} \right),\\
		\zeta^2 \frac{\overline{P}}{\sigma_{\widehat{\mathbf{v}}}^2} & = &  \lambda_{1}^2 \lambda_{2}^2  - 1.
		\end{array} \right. 
		\end{flalign}
		It can be verified that $\overline{\lambda}_{1}, \overline{\lambda}_{2} = \pm \lambda$, where $\overline{\lambda}$ denotes the real root of
		\begin{flalign} 
		\left( \frac{1}{1 + g_2 \lambda^{-2}} \right)^2 \frac{\overline{P}}{\sigma_{\widehat{\mathbf{v}}}^2}  = \lambda^4  - 1
		\end{flalign}
		with the largest absolute value.
		Correspondingly, the lower bound becomes
		\begin{flalign} \label{ma2}
		\log \left| \overline{\lambda}_{1} \overline{\lambda}_{2} \right| = \log \left| \overline{\lambda}^2 \right|.
		\end{flalign}
		
		This lower bound is indeed tight, i.e., \eqref{ma2} is the feedback capacity of the channel given in \eqref{machannel}; see discussions in \cite{kim2006feedback}. Meanwhile, \eqref{ma2} is also equivalent to
		\begin{flalign} \label{n2}
		\log \left| \overline{\rho} \right|,
		\end{flalign}
		where $\overline{\rho}$ denotes real root of
		\begin{flalign} 
		\left( \frac{1}{1 + g_2 \rho^{-1}} \right)^2 \frac{\overline{P}}{\sigma_{\widehat{\mathbf{v}}}^2}  = \rho^2  - 1
		\end{flalign}
		with the largest absolute value.
		As a matter of fact, \eqref{n2} is the same as the feedback capacity with the first-order MA noise \cite{kim2006feedback}
		\begin{flalign}
		\mathbf{v}_{k}  
		=    \widehat{\mathbf{v}}_k + g_2 \widehat{\mathbf{v}}_{k-1}, \nonumber
		\end{flalign}
		
	\end{itemize}

\end{example}

Meanwhile, \eqref{integrate6} and Fig.~\ref{kalman10} essentially provide a recursive coding scheme/algorithm to achieve the lower bound in Theorem~\ref{fc4}; see also discussions in Observation~\ref{observation1}. 

\begin{theorem}
	One class of recursive coding scheme to achieve the lower bound in Theorem~\ref{fc4} is given by
	\begin{flalign}
	\left\{ \begin{array}{rcl}
	\widetilde{\mathbf{x}}_{k+1}&=&A \widetilde{\mathbf{x}}_{k} +\mathbf{u}_k, \\
	\mathbf{y}'_{k}&= & C \widetilde{\mathbf{x}}_{k}, \\
	\mathbf{e}'_{k}&=& - \mathbf{y}'_k + \mathbf{v}_k, \\
	\mathbf{u}_{k}&=& \widehat{K} \left( \mathbf{e}'_{k} - \sum_{i=1}^{p} f_{i} \mathbf{e}'_{k-i} \right) - \sum_{j=1}^{q} g_{j} \mathbf{u}_{k-j},
	\end{array}
	\right.
	\end{flalign}
	where
	\begin{flalign} 
	A = T \left[
	\begin{array}{cccc}
	\overline{\lambda}_{1} & 0 & \cdots & 0\\
	0 & \overline{\lambda}_{2} & \cdots & 0\\
	\vdots & \vdots & \ddots & \vdots\\
	0 & 0 & \cdots & \overline{\lambda}_{n}
	\end{array}
	\right] T^{-1},
	\end{flalign}
	and
	\begin{flalign} 
	C = \left[
	\begin{array}{cccc}
	1 & 1 & \cdots & 1
	\end{array}
	\right] T^{-1},
	\end{flalign} 
	while $\widehat{K}$ can be found by \eqref{gain22}. Herein, $T \in \mathbb{R}^{n \times n}$ can be any invertible matrix. Note in particular that $- \mathbf{y}'_0$ represents the message (cf. discussions in \cite{kim2010feedback}). 
\end{theorem}

We next provide two examples on the recursive coding schemes to achieve the feedback capacity in Example~\ref{casethree} and Example~\ref{case}, respectively.

\begin{example}
One class of recursive coding scheme to achieve the feedback capacity in Example~\ref{casethree} is given by
\begin{flalign} \label{coding3}
\left\{ \begin{array}{rcl}
\widetilde{\mathbf{x}}_{k+1}&=&\overline{\lambda} \widetilde{\mathbf{x}}_{k} +\mathbf{u}_k, \\
\mathbf{y}'_{k}&= & c \widetilde{\mathbf{x}}_{k}, \\
\mathbf{e}'_{k}&=& - \mathbf{y}'_k + \mathbf{v}_k, \\
\mathbf{u}_{k}&=& \widehat{K} \left( \mathbf{e}'_{k} - f_{1} \mathbf{e}'_{k-1} \right) - g_{1} \mathbf{u}_{k-1},
\end{array}
\right.
\end{flalign}
where $c$ can be any non-zero real number.
Correspondingly, \eqref{feedbackcoding} reduces to
\begin{flalign} \label{arma3}
L \left( z \right) 
= \frac{ c \widehat{K} \left( 1 - f_{1} z^{-1} \right) }{ \left( z - \overline{\lambda} \right) \left( 1 + g_{1} z^{-1} \right)}
= \frac{ c \widehat{K} \left( z - f_{1}  \right) }{ \left( z - \overline{\lambda} \right) \left( z + g_{1} \right)},
\end{flalign}
while \eqref{gain22} becomes
\begin{flalign}
\widehat{K} 
= \frac{\left( \overline{\lambda}^2 -1\right) \left( 1 +  g_{1} \overline{\lambda}^{-1} \right) }{\overline{\lambda} \left( 1 - f_{1} \overline{\lambda}^{-1} \right)}
= \frac{\left( \overline{\lambda}^2 -1\right) \left( \overline{\lambda} +  g_{1} \right) }{\overline{\lambda} \left( \overline{\lambda} - f_{1} \right)},
\end{flalign} 
since in this case 
\begin{flalign}
P = \frac{\left( \overline{\lambda}^2 -1\right) \left( 1 +  g_{1} \overline{\lambda}^{-1} \right)^2 }{c \left( 1 - f_{1} \overline{\lambda}^{-1} \right)^2} \sigma_{\widehat{\mathbf{v}}}^2.
\end{flalign}
Furthermore, one might also compare \eqref{coding3} and \eqref{arma3} respectively with the coding scheme and $B \left( z \right)$ for the first-order ARMA noises as presented in \cite{kim2010feedback} by noting that
	\begin{flalign} 
B \left( z \right)
= \frac{- L \left( z \right)}{ 1 + L \left( z \right)}
= \frac{ - c \widehat{K} \left( z - f_{1}  \right) }{ \left( z - \overline{\lambda} \right) \left( z + g_{1} \right) + c \widehat{K} \left( z - f_{1}  \right)}.
\end{flalign}
\end{example}


\begin{example}
One class of recursive coding scheme to achieve the feedback capacity in Example~\ref{case} (as in Case~2 therein) is given by
\begin{flalign} \label{coding}
\left\{ \begin{array}{rcl}
\widetilde{\mathbf{x}}_{k+1}&=&A \widetilde{\mathbf{x}}_{k} +\mathbf{u}_k, \\
\mathbf{y}'_{k}&= & C \widetilde{\mathbf{x}}_{k}, \\
\mathbf{e}'_{k}&=& - \mathbf{y}'_k + \mathbf{v}_k, \\
\mathbf{u}_{k}&=& \widehat{K} \mathbf{e}'_{k}  - g_{2} \mathbf{u}_{k-2},
\end{array}
\right.
\end{flalign}
where 
\begin{flalign} 
A = \left[
\begin{array}{cc}
\overline{\lambda} & 0\\
0 & -\overline{\lambda}\\
\end{array} \right] T^{-1},~C = \left[
\begin{array}{cc}
1 & 1\\
\end{array} \right] T^{-1}.
\end{flalign}
where $T^{-1} \in \mathbb{R}^{2 \times 2}$ can be any invertible matrix.
Correspondingly, \eqref{feedbackcoding} reduces to
\begin{flalign} \label{arma}
L \left( z \right) = \frac{C \left( z I  - A  \right)^{-1} \widehat{K}}{ 1 + g_2 z^{-2} },
\end{flalign}
where $\widehat{K}$ can be found by \eqref{gain22}. 
\end{example}

\section{Conclusion}

In this paper, from the perspective of a variant of the Kalman filter, we have obtained explicit lower bounds on the feedback capacity of channels with  any finite-order ARMA Gaussian noises, as well as the accompanying recursive coding schemes to achieve them. 
Potential future research problems include investigating the tightness of the lower bounds beyond the cases considered in this paper. 

It is also worth mentioning that the results presented in this paper represent
the relatively preliminary explorations under the current
framework; particularly, we considered a very special class of the plant parameters $A$ and $C$ with a simple structure.
For future research, it might be interesting to investigate further the structure of the plant parameters, to make use of the yet unexploited degrees of freedom therein, 
which may potentially lead to tighter bounds and gain additional insights into the feedback capacity problem.

For another possible future research direction, examining the gap between the expressions of \cite{liu2018feedback} and ours might likely either simplify the expression in \cite{liu2018feedback} or point to structures of the plant parameters with tighter bounds in our approach.



%

%

%
%




\balance

\bibliographystyle{IEEEtran}
\bibliography{references}
\end{document}